\documentclass[11pt,journal]{IEEEtran}

\usepackage[utf8]{inputenc}
\usepackage[english]{babel}
\usepackage{epsfig}
\usepackage{amsfonts}
\usepackage{ifpdf}
\usepackage{multirow}
\usepackage{amssymb}
\usepackage{amsthm}
\usepackage{mathtools}
\usepackage{mathrsfs}
\usepackage{textgreek}
\usepackage{grffile}
\usepackage{xcolor}
\usepackage{bm}
\usepackage{cite}
\usepackage{setspace}
\usepackage{nccmath}
\usepackage{float}
\usepackage{mathtools}

\usepackage[usestackEOL]{stackengine}

  \stackMath

\ifCLASSINFOpdf
\else
\fi
\usepackage{amsmath}
\usepackage{algorithmic}
\usepackage{algorithm}
\usepackage{graphicx}

\newcommand{\norm}[1]{\left\lVert#1\right\rVert}

\ifCLASSOPTIONcompsoc
  \usepackage[caption=false,font=normalsize,labelfont=sf,textfont=sf]{subfig}
\else
  \usepackage[caption=false,font=footnotesize]{subfig}
\fi

\usepackage{stfloats}
\begin{document}
%

\title{Geolocation with Large LEO Constellations: Insights from Fisher Information}
%
\author{Don-Roberts~Emenonye,
        Harpreet~S.~Dhillon,
        and~R.~Michael~Buehrer

}
%
\markboth{}%
{Comminiello \MakeLowercase{\textit{et al.}}: Author Guidelines for Columns \& Forum Articles of IEEE SPM}


\maketitle
%
%
%
%
%
\section*{}
\label{sec:abstract}
Interest in the use of the low earth orbit (LEO) in space - from $160 \text{ km}$ to $2000 \text{ km}$ - has skyrocketed; this is evident by the fact that National Aeronautics and Space Administration (NASA) has partnered with various commercial platforms like Axiom Space, Blue Origin, SpaceX, Sierra Space, Starlab Space, ThinkOrbital, and Vast Space to deploy satellites. 
The most apparent advantage of satellites in LEO over satellites in Geostationary (GEO) and medium earth orbit (MEO) is their closeness to the earth; hence, signals from LEOs encounter lower propagation losses and reduced propagation delay, opening up the possibility of using these LEO satellites for localization. This article reviews the existing signal processing algorithms for localization using LEO satellites, introduces the basics of estimation theory, connects estimation theory to model identifiability with Fisher Information Matrix (FIM), and with the FIM, provides conditions that allow for $9$D localization of a terrestrial receiver using signals from multiple LEOs (unsynchronized in time and frequency) across multiple time slots and multiple receive antennas. We also compare the structure of the information available in LEO satellites with the structure of the information available in the Global Positioning System (GPS).
%
%
%
%
\section*{Introduction and Background}
\label{sec:scope}
\IEEEPARstart{F}{uturistic} use cases of wireless signals such as vehicle-vehicle communications, robot-robot interactions, and augmented reality will require highly accurate $3$D position, $3$D velocity, and $3$D orientation estimates; hence,  there is a need to investigate novel techniques for providing $9$D localization. Currently, the most ubiquitous method of providing positioning information is the Global Navigation Satellite Systems (GNSS). In strategic documents released by most countries, navigation is defined as driving in a secure and safe mode from a source to a destination. By definition, navigation is thus the determination of the positions of a receiver on a temporal scale in the same order as the speed of the receiver, and such determination is usually done by regularly observing electromagnetic signals. The hope is that the observations of the electromagnetic signal contain geometric information that can help determine the positions of the receiver node. In this case of using satellites for positioning, the geometric information extracted from the electromagnetic signals is the propagation times from the satellites to the receiver. While these propagation times and the positions of satellites with respect to a specific reference frame can be used to compute the position of a receiver, it should be noted that the lack of synchronization between the satellites and the receiver means that time differences are more often than not used to compute the receiver position. The type of measurements - propagation times or difference of propagation times- determines the positioning solution type. The position of the receiver is the intersection of three circles in $2$D or the intersection of three spheres in $3$D when the propagation times are used and is the intersection of hyperbolas circles in $2$D or the intersection of three hyperboloids in $3$D when the difference of propagation times. 

Although the global positioning system (GPS) \cite{mcneff2002global} is the most popular  GNSS constellation \cite{zekavat2011handbook}, there are five other constellations. They are Galileo \cite{henkel2008precise}, BeiDou Navigation Satellite System (BDS), GLONASS (Global Navigation Satellite System) \cite{zeng2015multiangle}, Indian Regional Navigation Satellite System (IRNSS) \cite{das2022indian} or Navigation Indian Constellation (NavIC), and Quasi-Zenith Satellite System (QZSS) \cite{wu2004performance}.

The GPS system, managed and controlled by the United States Air Force, is composed of three key components: i) 24 satellites in Medium Earth Orbit (MEO), ii) ground stations, and iii) a receiver entity \cite{mcneff2002global}. These satellites are positioned at an altitude of 11,900 km, spread across six orbital planes, each inclined at a $55^\circ$ angle, with four satellites per plane. Each GPS satellite is equipped with four rubidium clocks, but only one is active at any given time. The satellites are also equipped with an onboard computer that generates navigation messages. These messages contain critical data such as the satellite's position and precise time, which the receiver uses to estimate the time difference of arrival and calculate the correct time to within 10 nanoseconds.

GPS’s accuracy is extraordinary: it can estimate timing to within one-millionth of a second, velocity to within 0.45 m/s, and positioning to within 30 meters. Ground stations play an essential role by tracking satellite orbits and maintaining clock synchronization. These stations calculate the positions of the satellites, which are uploaded back to the satellites for orbit corrections. These corrections are effective for several weeks. Additionally, ground stations maintain the onboard clocks to nanosecond precision, in alignment with the global coordinated time (UTC) managed by the Master Clock at the U.S. Naval Observatory.

Galileo isn’t just another GNSS—it’s the European Union's powerhouse, designed by the European Space Agency and run by the European Union Agency for the Space Program. The system is broken down into three essential parts: i) the satellite constellation, ii) ground stations, and iii) the receiver entity. There are $30$ satellites in MEO, each of them broadcasting precise time signals, ephemeris data, and other critical information. These satellites sit at an altitude of $23,222 \text{ km}$, with a $56^{\circ}$ orbital inclination across three perfectly spaced orbital planes. Now, each satellite is armed with two top-tier hydrogen maser clocks. These atomic clocks measure time to within $0.45 \text{ ns}$ over $12 \text{ hours}$ using a hyper-stable $1.4 \text{ GHz}$ transition in a hydrogen atom. And just to be clear, there are two of these high-precision timekeepers. In case these clocks malfunction, the backup rubidium clock takes over, measuring the time to within $1.8 \text{ ns}$ over $12 \text{ hours}$. Again, two of these. So, there are four types of clocks in total, but only one of them is operating at any given time. Under normal operations, the primary master clock cranks out the reference frequency that powers the navigation signal. If this clock fails, the rubidium clock steps in seamlessly, and the other two clocks are ready to take over. If the problem lingers for days and it’s a maser clock failure, the second maser clock will step up, with the rubidium clock going into standby mode. So, no matter what happens, Galileo satellites always guarantee a navigation signal—thanks to those four clocks. Meanwhile, ground stations are always in operation, tracking satellite positions and performing periodic orbit corrections to keep the system firing on all cylinders.

The BeiDou Navigation Satellite System (BDS) was strategically engineered through a phased approach. Phase one, termed BDS-1, initiated in 1994 and deployed in 2000, delivered fundamental positioning, timing, and short-message communication exclusively to Chinese users. Building upon this, BDS-2, fully compatible with BDS-1, commenced development in 2004 and became operational in 2012. This phase introduced passive positioning capabilities, significantly enhancing location accuracy, velocity estimation, timing precision, and messaging services across the Asia-Pacific region. The most advanced iteration, BDS-3, launched in 2020, expanded upon BDS-2, offering a global suite of functionalities, including high-precision positioning, navigation, and timing (PNT), global short-message communication, and international search and rescue (SAR). Additionally, BDS-3 delivers specialized services such as satellite-based augmentation, ground augmentation, precise point positioning (PPP), and regional short-message communication, specifically tailored for users in China and adjacent regions. BDS-3 maintains a formidable 30-satellite constellation, comprising 3 Geostationary Orbit (GEO) satellites, 3 Inclined Geosynchronous Orbit (IGSO) satellites, and 24 Medium Earth Orbit (MEO) satellites. The GEO satellites, stationed at $35,786 \text{ km}$, are positioned at $80^{\circ} \text{ E}$, $110.5^{\circ} \text{ E}$, and $140^{\circ} \text{ E}$, respectively. The IGSO satellites, also at $35,786 \text{ km}$, maintain an orbital inclination of $55^{\circ}$ relative to the equatorial plane. The MEO satellites, forming a Walker24/3/1 constellation, operate at $21,528 \text{ km}$ with an identical $55^{\circ}$ inclination. For unparalleled timing accuracy, BDS-3 satellites incorporate Passive Hydrogen Masers (PHMs) as the primary atomic clock and Rubidium Atomic Frequency Standards (RAFSs) as secondary backups. The entire BDS infrastructure is supported by a ground control segment for system monitoring and maintenance and a user segment facilitating seamless access to BDS services.

The Russian Federation's GLONASS, originally deployed by the Soviet Union in the $1970\text{s}$, has undergone four major iterations: Glonass (1982–2005), Glonass-M (2003–2016), Glonass-K (2011–2018), and the ongoing Glonass-K2 (2017–present). Structurally, GLONASS consists of three key subsystems: i) The GLONASS Space Vehicle Subsystem, comprising $24$ active navigation satellites; ii) The Command and Control System, responsible for satellite management via a network of ground-based facilities; iii) The User Equipment (UE) Subsystem, which includes GLONASS-compatible receivers. The constellation of $24$ satellites operates in three orbital planes at an altitude of $19140 \text{ km}$ (MEO), inclined at $64.8^{\circ}$ relative to the equatorial plane. The command and control system integrates critical infrastructure, including the system control center, central synchronizer, tracking and telemetry stations, and laser ranging facilities. The system control center orchestrates satellite mission control, generates navigation messages, and executes essential operational uploads to ensure mission reliability, while also monitoring and responding to satellite anomalies. Precision in GLONASS timing is anchored in cesium atomic clocks, delivering timekeeping accuracy within an astonishing $10^{-12} \text{ to } 10^{-14} \text{ s}$, ensuring robust synchronization for global navigation applications.

India’s IRNSS, engineered and maintained by the government, is structured into three core segments: the space segment, the ground segment, and the user segment. The space segment comprises a constellation of $7$ satellites—three stationed in Geostationary Orbit (GEO) and four in Geosynchronous Orbit (GSO). The GEO satellites are positioned at $32.5^{\circ} \text{ E}$, $83^{\circ} \text{ E}$, and $131^{\circ} \text{ E}$, while the GSOs intersect the equator at $55^{\circ} \text{ E}$ and $111.75^{\circ} \text{ E}$. Each satellite, designated IRNSS-1A through IRNSS-1G, was sequentially deployed: IRNSS-1A on July 1, 2013, IRNSS-1B on April 4, 2014, IRNSS-1C on October 15, 2014, IRNSS-1D on March 28, 2015, IRNSS-1E on January 20, 2016, IRNSS-1F on March 10, 2016, and IRNSS-1G on April 28, 2016. The ground segment serves as the backbone of IRNSS, ensuring operational integrity and performance optimization. It encompasses:
i) The IRNSS spacecraft control facility for satellite operations,
ii) The navigation center for signal processing and system management,
iii) IRNSS range and integrity monitoring stations to validate signal accuracy,
iv) The IRNSS network timing center for synchronization,
v) IRNSS ranging stations for precise satellite positioning,
vi) Laser ranging stations for high-precision orbit determination, and
vii) The IRNSS data communication network, facilitating seamless data exchange across all system components. 

Japan's Quasi-Zenith Satellite System (QZSS) is engineered to enhance GPS functionality, particularly in Japan’s dense urban landscapes, by boosting positioning accuracy and reliability. The system is structured into three segments: space, ground, and user. The space segment consists of four satellites—three deployed in highly elliptical orbits (HEO) to ensure optimal coverage over Japan, and one stationed in geostationary orbit (GEO) for continuous signal broadcasting. The inaugural satellite, QZS-1, was launched on September 11, 2010, with the remaining three following in 2017. The ground segment is responsible for maintaining and controlling the constellation. It includes a master control station that generates navigation messages, tracking control stations for satellite monitoring, laser ranging stations for high-precision orbit determination, and monitoring stations to ensure system integrity and performance.

{\em Although these GNSS systems are usually reliable for navigation services, they may not be trustworthy or accurate in deep urban canyons or during GNSS spoofing. Hence, there is a need to provide an alternative that either replaces these GNSS systems or augments them. Inspired by this, in this paper, we present LEO satellites as a means of replacing or supplementing the current GNSS systems. We first present a tutorial-like summary of the signal processing algorithms in the existing LEO-based localization literature. Subsequently, we introduce the basics of estimation theory and link this theory to FIM-based model identifiability. We then use the FIM to provide an investigation of the number of unsynchronized LEO satellites, observation time slots, and the number of receive antennas that allow for different levels of $9$D localization.  We also compare the structure of the information available in LEO satellites with the structure
of the information available in the Global Positioning
System (GPS).}

\section*{State-of-the-art LEO Satellite Constellations for Localization }
The research on LEO satellites can be categorized into two main approaches: dedicated studies \cite{Fundamentals_of_LEO_Based_Localization,Joint_9D_Receiver_Localization,emenonye2023_MILCOM_conf_9D_localization,emenonye2023_MILCOM_conf_9D_localization_1,emenonye2023_VTC_conf_Minimal,emenonye2023_VTC_conf_unsyn,Fundamental_Performance_Bounds_for_Carrier_Phase_Positioning_LEO_PNT,Broadband_LEO_Constellations_for_Navigation,Economical_Fused_LEO_GNSS,Empowering_the_Tracking_Performance_of_LEOBased_Positioning_by_Means_of_Meta_Signals,Performance_Analysis_of_a_Multi_Slope_Chirp_Spread_Spectrum,Integrated_Communications_and_Localization_for_Massive_MIMO_LEO_Satellite,dureppagari2023ntn,dureppagari2024leo} and opportunistic studies \cite{Psiaki2020NavigationUC,Kassas2019NewAgeSN,Navigation_With_Differential_Carrier_Phase_Measurements_From_Megaconstellation_LEO_Satellites,A_Hybrid_Analytical_Machine_Learning_Approach_for_LEO_Satellite_Orbit_Prediction,Doppler_effect_Downlink_Receivers_OFDM_Low_earth_orbit_satellites_Bandwidth_Synchronization_Doppler_positioning_low_Earth_orbit,Ad_Astra_STAN_With_Megaconstellation_LEO_Satellites,A_Hybrid_Analytical_Machine_Learning_Approach_for_LEO_Satellite_Orbit_Prediction_1,Assessing_Machine_Learning_for_LEO_Satellite_Orbit_Determination_in_Simultaneous_Tracking_and_Navigation,Cognitive_Navigation_With_Unknown_OFDM_signals_With_Application_Terrestrial_5G_Starlink,Observability_Analysis_of_Receiver_Localization_Pseudorange,Positioning_with_Starlink_LEO_Satellites_A_Blind_Doppler_Spectral_Approach,Receiver_Design_for_Doppler_Positioning_with_Leo_Satellites,Unveiling_Starlink_LEO_Satellite_OFDM_Like_Signal_Structure_Enabling_Precise_Positioning}. The difference between these two systems lies in the knowledge of the signal structure.

In \cite{Fundamentals_of_LEO_Based_Localization}, we present a rigorous look at $9$D localization. In that work, a system model that captures i) the delay from distinct satellites to a terrestrial receiver, ii) the changing Doppler, iii) the time offset between the LEO satellites, and iv) the frequency offset between the LEO satellites is developed. The FIM for these parameters is rigorously developed. Subsequently, the FIM for $3$D position, $3$D velocity estimation, and $3$D orientation are developed. These FIMs are used to develop the achievable accuracy for $6$D localization and $9$D localization. The work in \cite{Fundamentals_of_LEO_Based_Localization} is extended in \cite{Joint_9D_Receiver_Localization,emenonye2023_MILCOM_conf_9D_localization} to account for ephemeris uncertainty and the potential of utilizing known $5$G reference signals to both perform $9$D localization and estimate the LEO position and velocity. By definition, the channel model has three possible links in \cite{Joint_9D_Receiver_Localization}. These links are the LEO satellite-to-receiver link, the LEO satellite-to-base station link, and the base station-to-receiver link. The channel parameters in these links are studied using the FIM, giving us a unique perspective on the fundamental difference between these links. These FIMs are transformed to the location parameters, and essential insights are drawn. These key findings are as follows:   i) When utilizing a single LEO satellite, achieving accurate estimation of the $9$D location parameters and correcting the LEO’s position and velocity requires three base stations (BSs) and three time slots.   ii) With two LEO satellites, the same task—estimating the $9$D location parameters and refining the LEO’s position and velocity—also necessitates three BSs and three time slots.   iii) When employing three LEO satellites, the process of estimating the $9$D location parameters and adjusting the LEO’s position and velocity demands three BSs but requires four time slots instead of three.The question: ``Is $9$D localization possible with unsynchronized LEO satellites?" is answered in \cite{emenonye2023_MILCOM_conf_9D_localization_1}. The minimal configurations needed to provide $3$D positioning are presented in \cite{emenonye2023_VTC_conf_Minimal}, and a fundamental look at receiver orientation error correction using LEO satellites is presented in \cite{emenonye2023_VTC_conf_unsyn}. The authors in \cite{Fundamental_Performance_Bounds_for_Carrier_Phase_Positioning_LEO_PNT} develop fundamental bounds for delay-based positioning that account for integer ambiguities. This bound is evaluated as a function of carrier frequency, bandwidth, transmission power, and the number of base stations. A holistic view of LEO satellites with multiple antennas is presented for integrated sensing and communication in \cite{Integrated_Communications_and_Localization_for_Massive_MIMO_LEO_Satellite}. Localization bounds were also developed, and the effect on various precoding and beamforming strategies was studied.

The author in \cite{Psiaki2020NavigationUC} observes that LEO satellites move faster than the satellites in other GNSS; hence, there is more information in the observed Dopplers that are useful for $3$D positioning. The author defines three system models that can be used for a Doppler-based LEO positioning system. The algorithm uses eight satellites in a point-solution nonlinear least-squares batch filter to provide $3$D position estimation, $3$D velocity estimation, and clock parameters estimation (offset and offset rate). Accuracy for positioning is on the order of $1-5$ meters, absolute velocity estimation accuracy on the order of $0.01 \text{ to } 0.05 \text{ m/s}$, and clock offset accuracy on the order of $0.0001 \text{ to } 0.0010 \text{ s}$ 

Authors in \cite{Kassas2019NewAgeSN} remark that terrestrial receivers usually have an Inertial navigation system (INS) and a GPS receiver. The location estimation done by the INS is accurate in the short term. It drifts rapidly, and the estimates it provides deteriorate rapidly. In this sense, the GPS receiver complements the INS by providing stable location estimates. However, GPS signals are usually weak and are susceptible to jamming or interference. Due to these challenges, \cite{Kassas2019NewAgeSN} provides an investigation of INS combined with LEO satellites to provide navigation for a ground receiver and unmanned aerial vehicle (UAV). The authors extract pseudorange, Doppler, and pseudorange rate measurements and utilize these measurements to perform receiver $3$D positioning, LEO orbital correction, and LEO clock correction. An algorithm that depends on a Kalman filter is defined and validated using an experimental setup with Orbcomm satellites trying to localize a terrestrial receiver. Results indicate an error of $1419 \text{ m}$ and $419 \text{ m}$ is achievable without INS and with INS, respectively. 

In \cite{Navigation_With_Differential_Carrier_Phase_Measurements_From_Megaconstellation_LEO_Satellites}, a $3$D positioning algorithm is developed that takes into consideration the angles of elevation and azimuth. The algorithm is evaluated to show a positioning error of $14.8 \text{ m}$ while tracking a UAV over a distance of $12 \text{ km}$ for $2$ minutes using signals from only two Orbcomm. The authors in \cite{A_Hybrid_Analytical_Machine_Learning_Approach_for_LEO_Satellite_Orbit_Prediction} investigates the intersection of machine learning and LEO-based navigation. The work has two passes; in the first pass, a receiver with knowledge of its position uses an extended Kalman filter to track an Orbcomm satellite. In the second pass, the Orbcomm satellites with perfectly known Ephemeris localize a second receiver. The problem of satellite acquisition is tackled in \cite{Doppler_effect_Downlink_Receivers_OFDM_Low_earth_orbit_satellites_Bandwidth_Synchronization_Doppler_positioning_low_Earth_orbit}, an algorithm for tracking the fast-changing Doppler is employed, and Doppler measurements from six Starlink satellites are used to achieve a $2$D positioning error of $10 \text{ m}$ and $3$D positioning error of $22 \text{ m}$. The authors in \cite{Ad_Astra_STAN_With_Megaconstellation_LEO_Satellites} develop a framework named Simultaneous Tracking and Navigation (STAN) that estimates the LEO position as well as the receiver position using pseudorange and Doppler measurements. This STAN framework has two modes: i) tracking Mode in which GNSS measurements are available. To produce an estimate of the vehicle's position while tracking the LEO satellites, the GNSS and LEO measurements are fused in
a filter to aid the INS, and ii) STAN Mode in which GNSS measurements are unavailable, so the measurements from the LEO satellites are fused with INS to localize the receiver.

In \cite{A_Hybrid_Analytical_Machine_Learning_Approach_for_LEO_Satellite_Orbit_Prediction_1}, a machine learning framework with three stages is developed. In the first stage, a receiver that knows its position is used to observe LEO satellites and extract appropriate measurements. These measurements propagate a two-line element data file containing information about the LEO satellite position. In the second stage, the neural network is trained on
the estimated ephemeris and is used to propagate the LEO
satellite orbit for the period where the satellite is not in view. 
During the third stage, when the LEO satellite is not in view, a receiver without knowledge of its position uses the machine learning predicted LEO ephemeris and its carrier phase measurements from received LEO signals to estimate its position via an extended Kalman filter. An experimental result indicates that signals from an Orbcomm satellite achieve a position error of $448 \text{ m}$.

In \cite{Assessing_Machine_Learning_for_LEO_Satellite_Orbit_Determination_in_Simultaneous_Tracking_and_Navigation}, the following two frameworks are developed:  i) Kalman filter-based STAN and ii) Kalman filer-based STAN coupled with a time delay neural network. The first framework has a positioning accuracy of $10.6 \text{ m}$ while two LEO satellites are tracked with errors of  $71 \text{ m}$ and $26 \text{ m}$, respectively. The authors in \cite{Cognitive_Navigation_With_Unknown_OFDM_signals_With_Application_Terrestrial_5G_Starlink} propose an intelligent blind receiver for detecting unknown reference signals in slow and fast Doppler rate scenarios. The deciphered reference signals are passed through tracking loops for refinements. Subsequently, localization is performed for a terrestrial receiver by exploiting Starlink signals, and a positioning error of $6.5 \text{ m}$ is achieved. The question of localizing a receiver with a single Starlink or Orbcomm satellite is answered in \cite{Observability_Analysis_of_Receiver_Localization_Pseudorange}. In \cite{Positioning_with_Starlink_LEO_Satellites_A_Blind_Doppler_Spectral_Approach}, an opportunistic framework for utilizing Doppler measurements observed from Starlink satellites for positioning is developed. The reference signals are not known to the receiver. In that work, an analytical derivation of the received signal frequency spectrum is presented. This characterization accounts for the channel's fast-changing nature between the LEO satellite and a ground-based receiver. An algorithm based on the Kalman filter is used for Doppler tracking. Experiments indicate that the Doppler from the six Starlink LEO satellites can be tracked for 800 seconds with very tight accuracy. The Doppler measurements were used in a nonlinear least squares estimator, starting with an initial estimate $200 \text{ km}$ away from the actual position. The proposed approach achieves a final $4.3 \text{ m}$ accuracy.

The authors in \cite{Unveiling_Starlink_LEO_Satellite_OFDM_Like_Signal_Structure_Enabling_Precise_Positioning} present a framework for determining the structure of Starlink reference signals. The spectrum of these reference signals is given, and a frame length is established. A receiver that performs signal acquisition from multiple satellites estimates the reference signals is presented. A blind location algorithm is presented, and a positioning error of $6.5 \text{ m}$ is achieved with six satellites. { \em In the current paper, we bring together the findings in these various contributions to both summarize those findings, but more importantly to provide insight into the opportunities and limitations to using large LEO satellite constellations for localization.}
\section*{Mathematical Preliminaries}
LEO satellites are only helpful for localization if geometric information is in the signals collected at a receiver. To quantify the geometric information in the signals collected at a receiver from the LEO satellites, we parameterize the unknown geometric information as $\bm{\eta}$ and look to estimation theory. From estimation theory, the first step is to determine the error in estimating $\bm{\eta}$ with an estimator, $\hat{\bm{\eta}}$. The chosen error metric is the mean square error defined as
$$
\text{MSE} = \mathbb{E}_{}\Bigg[ (\hat{\bm{\eta}} - \bm{\eta})^2\Bigg],
$$
and the MSE can be expanded as 
$$
\text{MSE} = \mathbb{E}_{}\Bigg[ ((\hat{\bm{\eta}} - \mathbb{E}( \hat{\bm{\eta}})) + (\mathbb{E}( \hat{\bm{\eta}}) - {\bm{\eta}}))^2\Bigg].
$$
From the above expression, it is trivial to obtain an expression of the MSE in terms of the variance of the estimator and the bias associated with the estimator, as shown below
$$
\text{MSE} = \text{var}(\hat{\bm{\eta}}) + \text{b}^2(\hat{\bm{\eta}}).
$$
The bias defined as $\text{b}(\hat{\bm{\eta}}) = \mathbb{E}( \hat{\bm{\eta}}) - {\bm{\eta}}$ specifies 
the difference between the average estimate and the true value of the parameter. The next step is to design an estimator. An obvious design strategy is to design an estimator that minimizes the MSE. However, this strategy has a flaw: the MSE is a function of the unknown parameter. Hence, designing an estimator based on the MSE can be non-trivial since the design process depends on the unknown parameter. Another strategy is to focus on designing an estimator that minimizes the variance
$$
\text{var}(\hat{\bm{\eta}}) = \mathbb{E}_{}\Bigg[ (\hat{\bm{\eta}} - \mathbb{E}_{}(\hat{\bm{\eta}}))^2\Bigg].
$$
With this strategy, the estimator does not depend on the unknown parameter. It is important to note that this estimator is the same as the estimator that results from minimizing the MSE when the bias is zero. Hence, an estimate derived by minimizing the variance is the same as an unbiased estimate derived by minimizing the MSE. This special case of minimizing the MSE when the bias is zero results in the minimum variance unbiased (MVU) estimator. This MVU has the advantage of concentrating the probability density function (PDF) of the error around zero, and this reduces the probability of having substantial error values. This discussion naturally leads to the question of the existence of the MVU estimator. From estimation theory, it is well known that the MVU estimator may or may not exist, and even if it exists, we may never be able to find it. One way to find the MVU estimator is through the Cramer Rao Lower Bound (CRLB), discussed below.

\subsection*{Cramer Rao Lower Bound}
The CRLB presents a lower bound on the variance of unbiased estimators. This bound is useful, and if the variance of any estimator attains this bound, then this estimator must be the MVU estimator. Even though the utility of this bound is evident, issues still have to be considered. Two of these issues are: i) the Cramer-Rao bound might not exist, as is the case when we have a parameter vector that is non-identifiable, and ii) even if the bound exists, it might still be non-trivial to find an estimator that attains. It should be emphasized that while the utility of this bound lies in the fact that if an estimator attains it, that estimator must be the MVU, even if no estimator achieves the CRLB, the CRLB still provides a benchmark for evaluating all other estimators. Next, we present a mathematical description of the CRLB and introduce the FIM concept. Given received signals, $\bm{y}$, if the PDF  specified by $\chi(\bm{y}; \bm{\eta})$ satisfies the regularity condition
$$
\mathbb{E}_{}\left[\frac{\partial \ln \chi(\bm{y}; \bm{\eta})}{\partial \bm{\eta}}\right]=0, \quad \text { for all } \bm{\eta},
$$
then the covariance matrix of  any unbiased estimator $\hat{\bm{\eta}}$ satisfies
$$
\mathrm{C}_{\hat{\bm{\eta}}}-\mathbf{J}_{ \bm{\bm{y}}; \bm{\bm{\eta}}}^{-1} \succeq 0,
$$
where $\succeq$ represents the positive semi-definitiveness of a matrix. The entries in the FIM are given as
$$
[\mathbf{J}_{ \bm{\bm{y}}; \bm{\bm{\eta}}}]_{[i,j]}=-\mathbb{E}_{}\left[\frac{\partial^2 \ln \chi(\bm{y}; \bm{\eta})}{\partial [\bm{\eta}]_{[i]} \partial [\bm{\eta}]_{[j]}}\right],
$$
where the first derivatives are evaluated at the true value of $\bm{\eta}$ and the expectation is taken with respect to $\chi(\bm{y}; \bm{\eta})$. Also, the lower bound $\mathrm{C}_{\hat{\bm{\eta}}} = \mathbf{J}_{ \bm{\bm{y}}; \bm{\bm{\eta}}}^{-1}$  is attained by  an unbiased estimator if and only if
$$
\frac{\partial \ln \chi(\bm{y}; \bm{\eta})}{\partial \bm{\eta}}=\mathbf{J}_{ \bm{\bm{y}}; \bm{\bm{\eta}}}(\mathbf{g}(\mathbf{y})-\bm{\eta}).
$$
Here, the function $\mathbf{g}$ is a $p$-dimensional function, and the unbiased estimator that attains the lower bound is $\hat{\mathbf{\eta}}=\mathbf{g}(\mathbf{y})$, and its covariance matrix is $\mathbf{J}_{ \bm{\bm{y}}; \bm{\bm{\eta}}}^{-1}$. This {\em efficient} unbiased estimator is also the MVU estimator \cite{kay1993fundamentals}. Since the diagonals of a positive semidefinite matrix must be positive, we have
$$
[\mathrm{C}_{\hat{\bm{\eta}}}-\mathbf{J}_{ \bm{\bm{y}}; \bm{\bm{\eta}}}^{-1}]_{[i,i]} \geq 0.$$
With this, we can write the lower bound on a specific parameter in the parameter vector as
$$
\operatorname{var}\left([\hat{\eta}]_{[i]}\right)=\left[\mathbf{C}_{\hat{\eta}}\right]_{[i,i]} \geq\left[\mathbf{J}_{ \bm{\bm{y}}; \bm{\bm{\eta}}}^{-1} \right]_{[i,i]}
$$
This lower bound is the CRLB of a specific parameter in the parameter vector. It is important to note the following points:
\begin{itemize}
    \item the {\em iff} condition for attaining the CRLB is important because if $\hat{\mathbf{\eta}}=\mathbf{g}(\mathbf{y})$ is efficient then it is the MVU estimator.
    \item also, we can combine the {\em iff} condition and the regularity condition to show that if $\hat{\mathbf{\eta}}=\mathbf{g}(\mathbf{y})$ is efficient then it is the unbiased estimator.    
\end{itemize} 
\subsection*{Fisher Information Matrix}
In this work, we are primarily concerned with the FIM because if the FIM is positive definite, then an estimate, $\hat{\bm{\eta}}$ of $\bm{\eta}$ exists. We proceed by presenting the FIM for geometric channel parameters $\bm{\eta}$ and location parameters $\bm{\kappa}$. To do this, we consider that a set of reference signals $\bm{x}$ is transmitted and received by a multiple antenna receiver. The received signal $\bm{y}$ is a function of  $\bm{\eta}$,  $\bm{\kappa}$, and the transmit signal,  $\bm{x}$, i.e.,   $\bm{y} =  f(\bm{\eta}_{}, \bm{\kappa}_{}, \bm{x}_{})$. The FIM for the geometric channel parameters is defined by $\mathbf{J}_{ \bm{\bm{y}}; \bm{\bm{\eta}}}$ and it captures the available  information about $\mathbf{\eta}$    present in $\mathbf{y}$. The corresponding CRLB of $\mathbf{\eta}$ provides a lower bound on the error associated with an unbiased estimate, $\hat{\mathbf{\eta}}$.  It is obtained  by summing the diagonals of $\mathbf{J}_{ \mathbf{\mathbf{y}}; \mathbf{\eta}}^{-1}$. The FIM for the location parameters is defined by $\mathbf{J}_{ \bm{\bm{y}}; \bm{\bm{\kappa}}}$ and it captures the available information about $\mathbf{\kappa}$  present in $\mathbf{y}$. The following bijective transformation relates these two FIMs
$$
\mathbf{J}_{\mathbf{y};\mathbf{\kappa}}  = \mathbf{\Upsilon}_{\mathbf{\kappa}} \mathbf{J}_{ \mathbf{\mathbf{y}}; \mathbf{\eta}} \mathbf{\Upsilon}_{\mathbf{\kappa}}^{\mathrm{T}},
$$where the entries in $\mathbf{\Upsilon}_{\mathbf{\kappa}}$ presents the unknown linear relationship between $\bm{\eta}$ and $\bm{\kappa}$.
It is important to note that given a parameter vector, $ \bm{\eta}_{} = \left[\bm{\eta}_{1}^{\mathrm{T}}, \bm{\eta}_{2}^{\mathrm{T}}\right]^{\mathrm{T}}$, where $\bm{\eta}_{1}$ is the  parameter of interest and $\bm{\eta}_{2}$ is the  nuisance parameters, the  resultant FIM has the structure 

$$\mathbf{J}_{ \bm{\bm{y}}; \bm{\eta}}=\left[\begin{array}{cc}\mathbf{J}_{ \bm{\bm{y}}; \bm{\eta}_1}^{}  & \mathbf{J}_{ \bm{\bm{y}}; \bm{\eta}_1, \bm{\eta}_2}^{} \\ \mathbf{J}_{ \bm{\bm{y}}; \bm{\eta}_1, \bm{\eta}_2}^{\mathrm{T}} &\mathbf{J}_{ \bm{\bm{y}}; \bm{\eta}_2}^{}\end{array}\right]$$
The equivalent FIM (EFIM)  of  parameter ${\bm{\eta}_{1}}$ is given by 
$$\mathbf{J}_{ \bm{\bm{y}}; \bm{\eta}_1}^{\mathrm{e}} =\mathbf{J}_{ \bm{\bm{y}}; \bm{\eta}_1}^{} - \mathbf{J}_{ \bm{\bm{y}}; \bm{\eta}_1}^{nu} =\mathbf{J}_{ \bm{\bm{y}}; \bm{\eta}_1}^{}-
\mathbf{J}_{ \bm{\bm{y}}; \bm{\eta}_1, \bm{\eta}_2}^{} \mathbf{J}_{ \bm{\bm{y}}; \bm{\eta}_2}^{-1} \mathbf{J}_{ \bm{\bm{y}}; \bm{\eta}_1, \bm{\eta}_2}^{\mathrm{T}}.$$ This EFIM captures all the required information about the parameters of interest present in the FIM; as observed from the relation $(\mathbf{J}_{ \bm{\bm{y}}; \bm{\eta}_1}^{\mathrm{e}})^{-1} = [\mathbf{J}_{ \bm{\bm{y}}; \bm{\eta}}^{-1}]_{[1:n,1:n]}$. The corresponding CRLB of $\bm{\eta}_{1}$ provides a lower bound on the error associated with an unbiased estimate,  $\hat{\bm{\eta}}_{1}$. 
It's obtained  by summing the diagonals of $[\mathbf{J}_{ \mathbf{\mathbf{y}}; \bm{\eta}_1}^{\mathrm{e}}]^{-1}$.


\section*{System Model}
We analyze a scenario involving $N_B$ LEO satellites tracked by a receiver equipped with $N_U$ antennas. The receiver has prior knowledge of the reference signals from the LEOs, which are observed over $N_K$ time slots, each separated by $\Delta_t$. At the $k^{\text{th}}$ time slot, the position of the $b^{\text{th}}$ LEO satellite is denoted as $\bm{p}{b,k}$, referenced to a global origin. Similarly, the position of the $u^{\text{th}}$ receive antenna at the same time slot is given by $\bm{p}{u,k}$, also defined relative to the global origin. The centroid of the receiver is represented as $\bm{p}_{U,k}$. The relative position of the $u^{\text{th}}$ receive antenna with respect to the receiver’s centroid is denoted by $\bm{s}_{u}$ and can be expressed as $\bm{s}_{u} = \bm{Q}_{U}\Tilde{\bm{s}}_{u}$ where $\Tilde{\bm{s}}_{u}$ is aligned with the global reference axis, and the matrix $$\bm{Q}_{U} = \bm{Q}\left(\alpha_{U}, \psi_{U}, \varphi_{U}\right)$$ defines a $3$D rotation transformation \cite{lavalle2006planning}. The receiver’s orientation angles are encapsulated in the vector. The orientation angles of the receiver are vectorized as $\bm{\Phi}_{U} = \left[\alpha_{U}, \psi_{U}, \varphi_{U}\right]^{\mathrm{T}}$. The relation between $\bm{p}{u,k}$ and $\bm{s}{u}$ is given by
$$\bm{p}_{u,k} = \bm{p}_{U,k} + \bm{s}_{u}.$$
The centroid position $\bm{p}_{U,k}$, referenced to the $b^{\text{th}}$ LEO satellite, is expressed as $$\bm{p}_{U,k} = \bm{p}_{b,k} + d_{bU,k}\bm{\Delta}_{bU,k}$$where $d_{bU,k}$ represents the distance from $\bm{p}{b,k}$ to $\bm{p}{U,k}$, and $\bm{\Delta}_{bU,k}$ is the corresponding unit direction vector, defined as

$$\bm{\Delta}_{bU,k} = [\cos \phi_{bU,k} \sin \theta_{bU,k}, \sin \phi_{bU,k} \sin \theta_{bU,k}, \cos \theta_{bU,k}]^{\mathrm{T}}.$$ 
During the $k^{\text{th}}$ transmission time slot, $\phi_{bU,k}$ and $\theta_{bU,k}$ denote the azimuth and elevation angles from the $b^{\text{th}}$ LEO satellite to the receiver, respectively.

\begin{figure}
\centering
    \fbox{\includegraphics[clip, trim=9.5cm 9.2cm 10.5cm 3cm,width=0.9
    \linewidth]{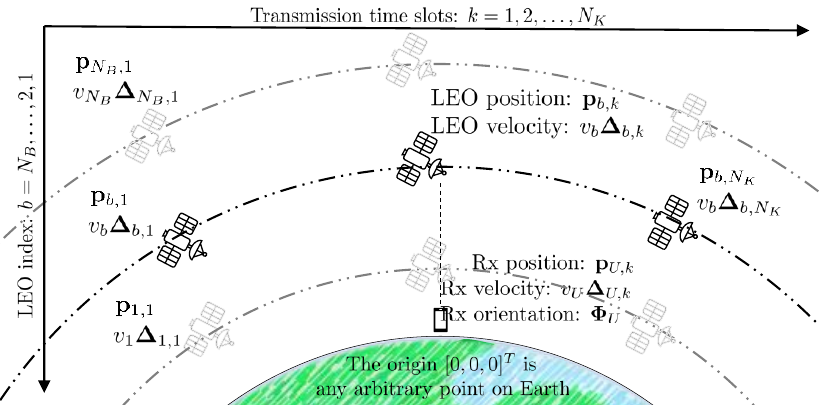}}
    \caption{LEO-based localization systems with $N_B$ LEOs transmitting during $N_K$ transmission time slots to a receiver with $N_U$ antennas.}
    \label{System_model_1}
\end{figure}

\subsection*{Transmit and Receive Processing}
At time $t$, the $b^{\text{th}}$ LEO satellite transmits using quadrature modulation, generating the following signal during the $k^{\text{th}}$ time slot
$$
        x_{b,k}[t] = s_{b,k}[t] \operatorname{exp}_{}{(j2 \pi f_c t )},
$$
the operating frequency is defined as $c = f_c \lambda$, where $s_{b,k}[t]$ represents the complex envelope of the signal transmitted by the $b^{\text{th}}$ LEO satellite during the $k^{\text{th}}$ time slot. Assuming a purely line-of-sight (LoS) channel, the received signal at the $u^{\text{th}}$ antenna for the $k^{\text{th}}$ time slot is given by
$$
\begin{aligned}
\label{equ:receive_signal}
y_{u,k}[t] &= \sum_{b}^{N_B} y_{bu,k}[t], \\ &= \sum_{b=1}^{N_b} \beta_{bu,k}  \sqrt{2} \Re\left\{s_{b,k}[t_{obu,k}]  \operatorname{exp}(j( 2 \pi f_{ob,k} t_{obu,k}))\right\} \\ &+ {n}_{u,k}[t], \\
&= {\mu}^{}_{u,k}[t] + {n}_{u,k}[t],
\end{aligned}
$$
where ${\mu}^{}{u,k}[t]$ represents the noise-free portion of the received signal, while ${n}{u,k}[t] \sim \mathcal{C}\mathcal{N}(0,N_0)$ models the thermal noise at the receiver's antenna array. The term $\beta_{bu,k}$ denotes the channel gain from the $b^{\text{th}}$ LEO satellite to the $u^{\text{th}}$ receive antenna during the $k^{\text{th}}$ time slot. The observed frequency at the receiver relative to the $b^{\text{th}}$ LEO is defined as
$$f_{ob,k} = f_c(1-\nu_{bU,k}) +\epsilon_{bU},$$ 
where $\nu_{bU,k}$ represents the Doppler shift induced by the relative motion between the satellite and receiver, and $\epsilon_{bU}$ accounts for any residual frequency offset.

The effective time duration of the received signal is given by
 $$t_{obu,k}= t-\tau_{bu,k} +\delta_{bU}$$ where $\delta_{bU}$ denotes the receiver’s time offset relative to the $b^{\text{th}}$ LEO satellite, and the signal propagation delay between the $u^{\text{th}}$ receive antenna and the $b^{\text{th}}$ LEO is expressed as
$$
\tau_{bu,k} \triangleq  \frac{\left\|\mathbf{p}_{u,k}-\mathbf{p}_{b,k}\right\|}{c}.
$$

The positions of the $b^{\text{th}}$ LEO satellite and the $u^{\text{th}}$ receive antenna at the $k^{\text{th}}$ time slot are given by
$$
\begin{aligned}
    \mathbf{p}_{b,k} &= \mathbf{p}_{b,0} + \Tilde{\mathbf{p}}_{b,k}, \\
    \mathbf{p}_{u,k} &= \mathbf{p}_{u,0} + \Tilde{\mathbf{p}}_{U,k},
\end{aligned}
$$
where $\mathbf{p}_{b,0}$ and $\mathbf{p}_{u,0}$ denote their respective reference positions, while $\Tilde{\mathbf{p}}_{b,k}$ and $\Tilde{\mathbf{p}}_{u,k}$ represent their traveled distances. These distances can be explicitly expressed as

$$
\begin{aligned}
\Tilde{\mathbf{p}}_{b,k} &= k \Delta_{t} v_{b} \mathbf{\Delta}_{b,k}, \\
    \Tilde{\mathbf{p}}_{U,k} &= k \Delta_{t} v_{U} \bm{\Delta}_{U,k},\\
    \end{aligned}
$$
where $v_{b}$ and $v_{U}$ are the velocities of the $b^{\text{th}}$ LEO satellite and the receiver, respectively. The corresponding unit direction vectors are given by

 $$\bm{\Delta}_{b,k} = [\cos \phi_{b,k} \sin \theta_{b,k}, \sin \phi_{b,k} \sin \theta_{b,k}, \cos \theta_{b,k}]^{\mathrm{T}},$$  $$\bm{\Delta}_{U,k} = [\cos \phi_{U,k} \sin \theta_{U,k}, \sin \phi_{U,k} \sin \theta_{U,k}, \cos \theta_{U,k}]^{\mathrm{T}}
 .$$ The velocity vectors of the LEO satellite and the receiver can thus be formulated as

$$
\boldsymbol{v}_b=v_b \boldsymbol{\Delta}_{b, k}, \quad \boldsymbol{v}_U=v_U \boldsymbol{\Delta}_{U, k}
$$

Consequently, the Doppler shift observed by the receiver from the $b^{th}$ LEO satellite is given by

$$
\nu_{b U, k}=\boldsymbol{\Delta}_{b U, k}^{\mathrm{T}} \frac{\left(\boldsymbol{v}_{b, k}-\boldsymbol{v}_{U, k}\right)}{c}.
$$

This formulation rigorously captures the interaction between satellite motion, signal propagation, and receiver dynamics, essential for precise signal modeling and system analysis.


\subsection*{Received Signal Properties}
To expose the properties of the received signal from the $b^{\text{th}}$ LEO satellites during the $k^{\text{th}}$ time slot, let's take a dive into the Fourier transform of the transmitted signal, defined as:

$$
S_{b,k}[f] \triangleq \frac{1}{\sqrt{2 \pi}} \int_{-\infty}^{\infty} s_{b,k}[t] \operatorname{exp}_{}{(-j2 \pi f t )} \; \;    d t.
$$
This mathematical tool uncovers the frequency domain characteristics of the signal, breaking it down into its spectral components. The received signal properties are
\subsubsection{Effective Baseband Bandwidth}
This  is all about the variance of the entire frequency range the signal occupies. From the system’s definition, we have:
$$
\alpha_{1b,k} \triangleq\left(\frac{\int_{-\infty}^{\infty} f^2\left|S_{b,k}[f]\right|^2 d f}{\int_{-\infty}^{\infty}\left|S_{b,k}[f]\right|^2 d f}\right)^{\frac{1}{2}}.
$$
It tells you exactly how spread out the signal’s power is across frequencies, giving you the signal’s frequency variance. It's a key measurement to understand the energy distribution, cutting through the noise and offering a hard-hitting analysis of signal behavior.
\subsubsection{Effective Bandwidth}
The effective bandwidth is
 $$\omega_{bU,k} = \Bigg[ \alpha_{1b,k}^2 +  2f_{ob,k} \alpha_{1b,k} \alpha_{2b,k} +  f_{ob,k}^2   \Bigg].$$
\subsubsection{Baseband-Carrier Correlation (BCC)}
This property is crucial for delivering a compact yet powerful representation of the mathematical description of the information packed into the received signals. It’s defined as $$
\alpha_{2b,k} \triangleq\frac{\int_{-\infty}^{\infty} f\left|S_{b,k}[f]\right|^2 d f}{\left(\int_{-\infty}^{\infty} f^2\left|S_{b,k}[f]\right|^2 d f \right)^{\frac{1}{2}} \left(\int_{-\infty}^{\infty}\left|S_{b,k}[f]\right|^2 d f\right)^{\frac{1}{2}}}.
$$
This equation zeroes in on the signal’s balance — it links the first moment of the signal’s frequency spectrum with its second moment and the total power. It’s like finding the sweet spot between how far the frequencies are spread and how concentrated the energy is, delivering a tight, insightful summary of the signal's characteristics.

\subsubsection{Root Mean Squared Time Duration}
The root mean squared (RMS) time duration for the signal traveling from the $b^{\text{th}}$ LEO satellite to the $u^{\text{th}}$ receive antenna in the $k^{\text{th}}$ time slot is rigorously characterized as

$$
\alpha_{obu,k} \triangleq\left(\frac{\int_{-\infty}^{\infty}  2 t_{obu,k}^{2} \left|s(t_{obu,k})\right|^2  \; dt_{obu,k}}{\int_{-\infty}^{\infty}  \left|s(t_{obu,k})\right|^2  \; dt_{obu,k}}\right)^{\frac{1}{2}}.
$$
This metric encapsulates the temporal dispersion of the received signal, quantifying the spread of its energy over time.

\subsubsection{Received Signal-to-Noise Ratio}
The Signal-to-Noise Ratio (SNR) is the critical metric that measures how strong the signal is relative to the noise across its entire frequency range. Formally, given the system model, the SNRs are defined as
$$
\underset{bu,k}{\operatorname{SNR}} \triangleq \frac{8 \pi^2 \left|\beta_{bu,k}\right|^2}{N_{01}} \int_{-\infty}^{\infty}\left|S_{b,k}[f]\right|^2 d f,
$$
If the channel gain remains unchanged across all receive antennas, we can express the SNR as:
$$
\underset{b,k}{\operatorname{SNR}} \triangleq \frac{8 \pi^2 \left|\beta_{b,k}\right|^2}{N_{01}} \int_{-\infty}^{\infty}\left|S_{b,k}[f]\right|^2 d f,
$$

When the same signal is transmitted across all $N_{K}$ time slots, and assuming the channel gain stays consistent across all receive antennas and time slots, the SNR is given by:

$$
\underset{b}{\operatorname{SNR}} \triangleq \frac{8 \pi^2 \left|\beta_{b}\right|^2}{N_{01}} \int_{-\infty}^{\infty}\left|S_{b}[f]\right|^2 d f,
$$

The analysis of these received signal properties plays a vital role in subsequent investigations concerning the use of LEO satellites for precise localization tasks. The SNR directly impacts the accuracy and reliability of positioning, making it indispensable in the quest to leverage LEOs for high-performance localization.

\section*{Available Information in the Received Signal}
In this section, we use the FIM to present the available information about the geometric channel parameters - delay, Doppler, channel gain, and the nuisance parameters - time and frequency offset. 
\subsection*{FIM for Channel Parameters}
\begin{figure}[H]
\centering
    {\includegraphics[clip,width=\linewidth]{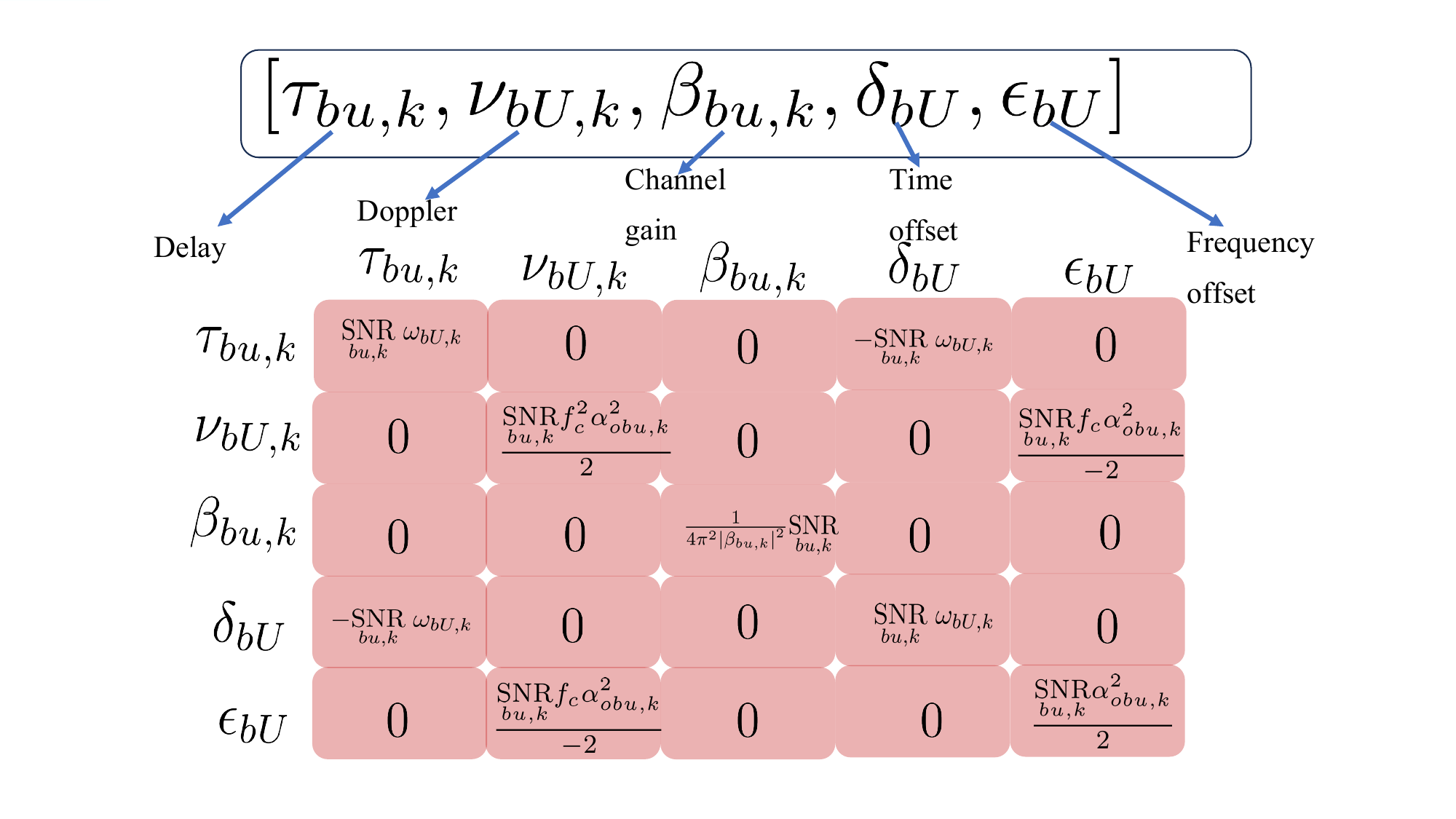}}
    \caption{Structure of the FIM considering the $b^{\text{th}}$ LEO Satellite, the $k^{\text{th}}$ time slot, and the $u^{\text{th}}$ receive Antenna.}
    \label{structure_of_FIM}
\end{figure}

Considering the $b^{\text{th}}$ LEO satellite, the channel parameters at the $u^{\text{th}}$ receive antenna during the $k^{\text{th}}$ time slot is given by Fig. \ref{structure_of_FIM}. 

This figure also describes the structure of the FIM considering the $b^{\text{th}}$ LEO satellite, the $k^{\text{th}}$ time slot, and the $u^{\text{th}}$ receive antenna. The term in the 1st row and 1st column indicates the information available in estimating the delay considering the $b^{\text{th}}$ LEO satellite, the $k^{\text{th}}$ time slot, and the $u^{\text{th}}$ receive antenna. This term, which is given by
$$
\underset{bu,k}{\operatorname{SNR}} \;  \omega_{bU,k}
$$
is simply the SNR  considering the $b^{\text{th}}$ LEO satellite, the $k^{\text{th}}$ time slot, and the $u^{\text{th}}$ receive antenna multiplied by the effective bandwidth. The other non-zero term in the first row or first column is 
$$
-\underset{bu,k}{\operatorname{SNR}} \;  \omega_{bU,k}.
$$
This term is the FIM that describes the interactions between the delay considering the $b^{\text{th}}$ LEO satellite, the $k^{\text{th}}$ time slot, and the $u^{\text{th}}$ receive antenna and the time offset. Notably, the zero entries in the first row or column indicate that the estimation accuracy achieved when estimating the delay is independent of the knowledge or lack of knowledge of the Doppler, channel gain, and frequency offset.

Considering the second row or column, there are only two non-zero terms. The first non-zero element represents the information about the Doppler considering the $b^{\text{th}}$ LEO satellite, the $k^{\text{th}}$ time slot, and the $u^{\text{th}}$ receive antenna. This term is
$$
\frac{ \underset{bu,k}{\operatorname{SNR}} f_{c}^2 \alpha_{obu,k}^2}{2}.
$$
This term is the SNR multiplied by the center frequency multiplied by the  root mean squared time duration. The other non-zero term relates the Doppler to the frequency offset and is 
$$
-\frac{ \underset{bu,k}{\operatorname{SNR}} f_{c}^2 \alpha_{obu,k}^2}{2}.
$$
The zero terms in the second row and column indicate that the estimation of the Doppler does not depend on the knowledge of the delays, channel gain, and time offset.
The third row or column in Fig. \ref{structure_of_FIM} indicates the available information about channel gain, and the only non-zero entry is
$$ \frac{1}{4 \pi^2 \left|\beta_{bu,k}\right|^2}\underset{bu,k}{\operatorname{SNR}}.$$
This entry specifies the information available to estimate the channel gain. The zero entries in this row or column indicate that the knowledge of delay, Doppler, time, and frequency offset does not affect the estimation accuracy of the channel gain. 
The fourth row or column presents the information available for estimating the time offset. The first term in the fourth row indicates the FIM relating the delay and the time offset and is
$$
-\underset{bu,k}{\operatorname{SNR}} \;  \omega_{bU,k}.
$$
While the second non-zero term in the fourth row is
$$
\underset{bu,k}{\operatorname{SNR}} \;  \omega_{bU,k}.
$$
This term is simply the SNR  considering the $b^{\text{th}}$ LEO satellite, the $k^{\text{th}}$ time slot, and the $u^{\text{th}}$ receive antenna multiplied by the effective bandwidth. The zero entries in this row indicate that only the Delay affects the estimation accuracy achievable when estimating the time offset. Out of all the entries in the fifth row or column, there are only two non-zero entries. The first non-zero entry is
$$
\frac{ \underset{bu,k}{\operatorname{SNR}} f_{c}\alpha_{obu,k}^2}{-2}.
$$
This term represents the relationship between the Doppler and frequency offset. The second non-zero entry is 
$$
\frac{ \underset{bu,k}{\operatorname{SNR}} \; \; \alpha_{obu,k}^2}{2}.
$$
The zero entries indicate that the delay, channel gain, and time offset do not affect the estimation accuracy achievable when estimating the frequency offset.

\subsection*{Transforming the Channel Parameters to the Location Parameters}
In the previous section, we presented the FIM for the channel parameters considering the $b^{\text{th}}$ LEO satellite, the $k^{\text{th}}$ time slot, and the $u^{\text{th}}$ receive antenna. To obtain the FIM for the location parameters, we vectorize the channel parameters in this section.  We can vectorize the delays received across all the antennas during the $k^{\text{th}}$ time slot as
$$
\bm{\tau}_{bU,k}
\triangleq\left[{\tau}_{b1,k}, {\tau}_{b2,k}, \cdots,
{\tau}_{bN_U,k}\right]^{\mathrm{T}}, 
$$
the next vectorization occurs considering the time slots and the $b^{\text{th}}$ LEO
$$
\bm{\tau}_{bU}
\triangleq\left[\bm{\tau}_{bU,1}^{\mathrm{T}}, \bm{\tau}_{bU,2}^{\mathrm{T}}, \cdots, \bm{\tau}_{bU,N_K}^{\mathrm{T}}\right]^{\mathrm{T}}.
$$
Focusing on the $b^{\text{th}}$ LEO satellite, the Doppler across all the $N_{K}$ transmission time slots is
$$
\bm{\nu}_{bU}
\triangleq\left[{\nu}_{bU,1}, {\nu}_{bU,2}, \cdots, \nu_{bU,N_K}\right]^{\mathrm{T}}.
$$
The channel gain in the LEO-receiver link can be placed in vector form as
$$
\bm{\beta}_{bU,k}
\triangleq\left[{\beta}_{b1,k}, {\beta}_{b2,k}, \cdots, {\beta}_{bN_U,k}\right]^{\mathrm{T}}, 
$$
and 
$$
\bm{\beta}_{bU}
\triangleq\left[\bm{\beta}_{bU,1}^{\mathrm{T}}, \bm{\beta}_{bU,2}^{\mathrm{T}}, \cdots, \bm{\beta}_{bU,N_K}^{\mathrm{T}}\right]^{\mathrm{T}}.
$$
If the channel gain stays constant across the $N_K$ time slots and $N_U$ receive antennas, it can be reduced to a single scalar $\beta_{bU}$. Consequently, the observable parameters from the $b^{\text{th}}$ LEO across the $N_U$ antennas over $N_K$ time slots can be forcefully structured into a vector form as:$$
\bm{\eta}_{bU} \triangleq\left[\bm{\tau}_{bU}^{\mathrm{T}}, \bm{\nu}_{bU}^{\mathrm{T}}, \bm{\beta}_{bU}^{\mathrm{T}}, \delta_{bU}, \epsilon_{bU}\right]^{\mathrm{T}}.
$$
We vectorize all the parameters across all $N_B$ LEO satellites as
        $$
    \bm{\eta} = \left[\bm{\eta}_{1U}^{\mathrm{T}},\cdots,\bm{\eta}_{N_BU}^{\mathrm{T}}\right]^{\mathrm{T}}.
    $$
    We can transform the FIM for $\bm{\eta}$ specified as $\mathbf{J}_{ \bm{\bm{y}}; \bm{\eta}}$ to the FIM for the location parameters specified as
    $\mathbf{J}_{ \bm{\bm{y}}; \bm{\kappa}}$. This transformation is carried out by the following
    $$\mathbf{J}_{\bm{y};\bm{\kappa}} \triangleq \mathbf{\Upsilon}_{\bm{\kappa}} \mathbf{J}_{ \bm{\bm{y}}; \bm{\eta}} \mathbf{\Upsilon}_{\bm{\kappa}}^{\mathrm{T}}. 
$$
The entries in $\mathbf{\Upsilon}_{\bm{\kappa}}$ is presented in \cite{Fundamentals_of_LEO_Based_Localization}.
\section*{FIM for Location Parameters}
To proceed in finding the available information for localization, we specify the location parameters 
$$
\begin{aligned}
    \bm{\kappa} = [\bm{p}_{U,0}, \bm{v}_{U,0}, \bm{\Phi}_{U},\bm{\zeta}_{1U},\cdots, \bm{\zeta}_{N_BU} ],
    \end{aligned}
    $$
$$\text{where}$$ 
$$
\begin{aligned}
\bm{\zeta}_{bU} &= \left[\bm{\beta}_{bU}^{\mathrm{T}}, \delta_{bU}, \epsilon_{bU}\right]^{\mathrm{T}}.
\end{aligned}
$$
With this parameter specification, we can specify the FIM for localization as $\mathbf{J}_{ \bm{\bm{y}}; \bm{\kappa}}$. However, we can divide the location parameter into the parameter of interest, $\bm{\kappa}_{1}$, and nuisance parameters, $\bm{\kappa}_{2}$. Here, $\bm{\kappa}_{1} = [\bm{p}_{U,0}, \bm{v}_{U,0}, \bm{\Phi}_{U}]$ and $\bm{\kappa}_{2} =   [\bm{\zeta}_{1U},\cdots, \bm{\zeta}_{N_BU}  ].$ Now, we can focus on the FIM for the parameters of interest using the EFIM, which is specified by $\mathbf{J}_{ \bm{\bm{y}}; \bm{\kappa}_{1}}^{\mathrm{e}}$. More rigorously, the EFIM for the location parameters is 
    $$\mathbf{J}_{ \bm{\bm{y}}; \bm{\kappa}_1}^{\mathrm{e}} =\mathbf{J}_{ \bm{\bm{y}}; \bm{\kappa}_1}^{} - \mathbf{J}_{ \bm{\bm{y}}; \bm{\kappa}_1}^{nu} .$$
 We define functions $
 \bm{F}_{\bm{v}}(\bm{w} ; \bm{x}, \bm{y}) \triangleq \mathbb{E}_{\bm{v}}\left\{\left[\nabla_{ \bm{x}} \ln f(\bm{w})\right]\left[\nabla_{\bm{y}} \ln f(\bm{w})\right]^{\mathrm{T}}\right\} 
 $ and $ \bm{G}_{\bm{v}}(\bm{w} ; \bm{x}, \bm{y})$. The function $ \bm{G}_{\bm{v}}(\bm{w} ; \bm{x}, \bm{y})$
 describes the loss of information in the FIM defined by $
 \bm{F}_{\bm{v}}(\bm{w} ; \bm{x}, \bm{y})$ due to uncertainty in the nuisance parameters. In other words, $
 \bm{F}_{\bm{v}}(\bm{w} ; \bm{x}, \bm{y})$ specifies the appropriate entries in $\mathbf{J}_{ \bm{\bm{y}}; \bm{\kappa}_1}^{}$ while $ \bm{G}_{\bm{v}}(\bm{w} ; \bm{x}, \bm{y})$ specifies the appropriate entries in  $\mathbf{J}_{ \bm{\bm{y}}; \bm{\kappa}_1}^{nu}$. In $\mathbf{J}_{ \bm{\bm{y}}; \bm{\kappa}_1}^{\mathrm{e}}$, there are nine $3 \times 3$ block matrices. Hence, in $ \bm{F}_{\bm{v}}(\bm{w} ; \bm{x}, \bm{y})$ and $ \bm{G}_{\bm{v}}(\bm{w} ; \bm{x}, \bm{y})$, there are also nine $3 \times 3$ block matrices. 

\subsection*{Available Information for $3$D Positioning and Comparison with GPS}
\begin{figure}[H]
\centering
        {\includegraphics[clip, trim=0.1cm 0.01cm .1cm 0.2cm,
    ,width=\linewidth]{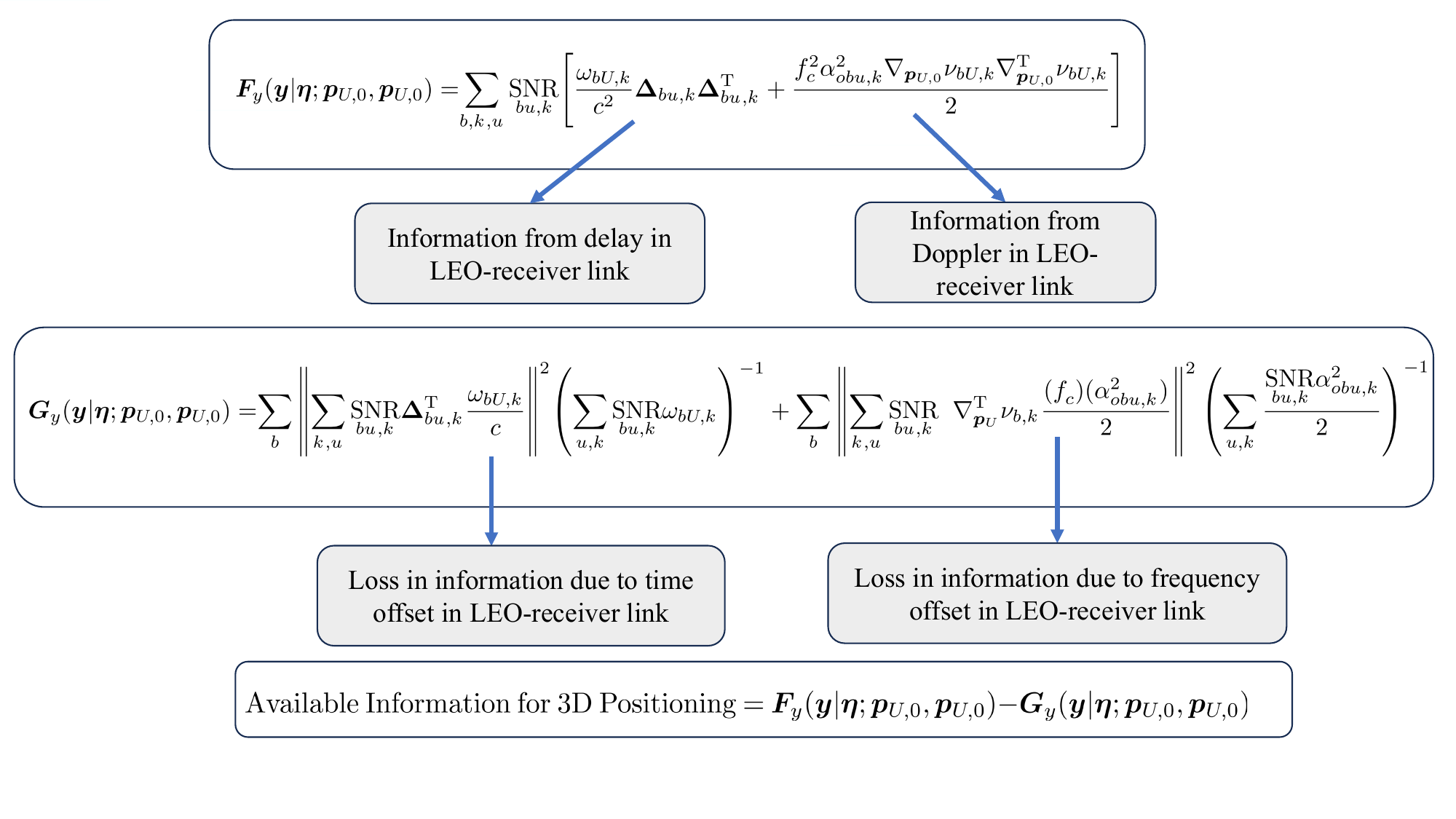}}
    \caption{The EFIM, $\mathbf{J}_{ \mathbf{\mathbf{y}}; \mathbf{\kappa}_1}^{\mathrm{e}}$, gives the available  information about $\mathbf{\kappa}_{1}$    present in $\mathbf{y}$. The entries in this EFIM, which are related to the $3$D position of the receiver, are given in this figure. This $3 \times 3$ matrix is located in the first block row and first block column in $\mathbf{J}_{ \mathbf{\mathbf{y}}; \mathbf{\kappa}_1}^{\mathrm{e}}$.}
    \label{EFIM_Position_3D}
\end{figure}

The FIM for $3$D positioning is the first block matrix in $\mathbf{J}_{ \bm{\bm{y}}; \bm{\kappa}_1}^{\mathrm{e}}$. This matrix is given in Fig. \ref{EFIM_Position_3D}. The information from the delays for the $3$D positioning is 
$$
{\sum_{b,k^{},u^{}} \underset{bu,k}{\operatorname{SNR}}  \frac{\omega_{bU,k} }{c^2}   \bm{\Delta}_{bu,k} 
 \bm{\Delta}_{bu^{},k^{}}^{\mathrm{T}}}.
$$
This shows that the information from delays depends on the SNR, the effective bandwidth, and the unit vector from the LEO satellites to the receive antennas. The summation is across the number of LEO satellites, the number of time slots, and the number of receive antennas. Hence, increasing these parameters strictly increases the available information for the $3$D positioning. {\em Although LEO and GPS satellites have different orbits (LEO vs. MEO), the structure of the information from the delays is identical when either system is used. The difference in the FIMs is captured by the SNR. The fact that LEO based systems are closer to the Earth's surface than GPS systems, means that their SNRs will be higher for the same transmit power. As we will discuss shortly, this is different when the Dopplers are used.} The information from the Dopplers for the $3$D positioning is 
$$
\sum_{b,k^{},u^{}} \underset{bu,k}{\operatorname{SNR}}  \frac{f_{c}^2 \alpha_{obu,k}^2\nabla_{\bm{p}_{U,0}} \nu_{bU,k} \nabla_{\bm{p}_{U,0}}^{\mathrm{T}}\nu_{bU,k}}{2}  
$$
where
$$
\nabla_{\bm{p}_{U,0}}\nu_{bU,k} \triangleq \frac{(\bm{v}_{b,k} - \bm{v}_{U,k}) - \bm{\Delta}_{bU,k}^{\mathrm{T}}(\bm{v}_{b,k} - \bm{v}_{U,k})\bm{\Delta}_{bU,k}}{c\bm{d}_{bU,k}^{-1}}.
$$
From this, we notice that FIM for Dopplers increases with SNR and transmit frequency.  {\em It is also evident that the information from Doppler for positioning decreases quadratically with the distance from the satellites. Hence, there is more information for Doppler-based positioning using LEO satellites than GPS satellites. This is precisely because of the distance of both orbits from the receiver. This explains why GPS satellites only use delay measurements for positioning, but LEO satellites can use both delays and Doppler measurements.} 
The loss in information due to the time offset between the $N_B$  LEO satellites and the receiver is 
$$
{\sum_{b}   \norm{\sum_{k^{},u^{}} \underset{bu^{},k^{}}{\operatorname{SNR}}\bm{\Delta}_{bu^{},k^{}}^{\mathrm{T}} \frac{ \omega_{bU,k}}{c} }}^{2}    {   \left(\sum_{u,k} \underset{bu,k}{\operatorname{SNR}} \omega_{bU,k}\right)^{\mathrm{-1}}  }.
$$
Notice that the information lost increases linearly with the number of LEO satellites. This is because the LEO satellites are unsynchronized with each other and the receiver. {\em This is a fundamental difference between LEO satellites and GPS; for GPS, this summation across satellites would scale much slower because the GPS satellites have an atomic clock and are synchronized with each other but not with the receiver.} A similar comparison can be made for the effect of frequency offset when LEO satellites are used vs. when GPS systems are used for $3$D positioning (see \ref{freq_offset_loss_information_position}).
\begin{equation}
\label{time_offset_loss_information_position}
    \begin{aligned}
&\underbrace{{\sum_{b}   \norm{\sum_{k^{},u^{}} \underset{bu^{},k^{}}{\operatorname{SNR}}\bm{\Delta}_{bu^{},k^{}}^{\mathrm{T}} \frac{ \omega_{bU,k}}{c} }}^{2}    {   \left(\sum_{u,k} \underset{bu,k}{\operatorname{SNR}} \omega_{bU,k}\right)^{\mathrm{-1}}  }}_\text{loss in information due to lack of time synchronization among LEO satellites}   \succeq
\\ &\underbrace{{  \norm{\sum_{b,k^{},u^{}} \underset{bu^{},k^{}}{\operatorname{SNR}}\bm{\Delta}_{bu^{},k^{}}^{\mathrm{T}} \frac{ \omega_{bU,k}}{c} }}^{2}    {   \left(\sum_{b,u,k} \underset{bu,k}{\operatorname{SNR}} \omega_{bU,k}\right)^{\mathrm{-1}}  }}_\text{loss in information due to lack of time syn. between GPS satellites and the receiver} 
\end{aligned}
\end{equation}
Equations \ref{time_offset_loss_information_position} and \ref{freq_offset_loss_information_position} shows the difference in the loss of information due to time and frequency offset when LEO satellites are used compared to GPS-based systems. 
\begin{equation}
\label{freq_offset_loss_information_position}
    \begin{aligned}
&\underbrace{\medmath{\sum_{b}\norm{{\sum_{k^{},u^{} } \underset{bu^{},k^{}}{\operatorname{SNR}} \; \; \nabla_{\bm{p}_{U}}^{\mathrm{T}} \nu_{b,k^{}}   }  \frac{(f_{c}^{}) (\alpha_{obu,k^{}}^{2})}{2} }^{2}  \left(\sum_{u,k} \frac{\underset{bu,k}{\operatorname{SNR}}  \alpha_{obu,k}^{2}}{2}\right)^{-1}}
}_\text{loss in information due to lack of freq. synchronization among LEO satellites}   \succeq
\\ &\underbrace{\medmath{\norm{{\sum_{b,k^{},u^{} } \underset{bu^{},k^{}}{\operatorname{SNR}} \; \; \nabla_{\bm{p}_{U}}^{\mathrm{T}} \nu_{b,k^{}}   }  \frac{(f_{c}^{}) (\alpha_{obu,k^{}}^{2})}{2} }^{2}  \left(\sum_{b,u,k} \frac{\underset{bu,k}{\operatorname{SNR}}  \alpha_{obu,k}^{2}}{2}\right)^{-1}}
}_\text{loss in information due to lack of freq. syn. between GPS satellites and the receiver} 
\end{aligned}
\end{equation}

The minimal configurations that allow for $3$D positioning are presented next. These configurations are the minimal configurations that allow the FIM in Fig. \ref{EFIM_Position_3D} to be positive definite. With a single time slot and no time or frequency offset, the following configurations allow for $3$D positioning:
\begin{itemize}
    \item one LEO satellites and multiple antennas.
    \item two LEO satellites and a single antenna.
\end{itemize}
With a single time slot and no time synchronization across the LEO satellites and the receiver, the following configurations allow for $3$D positioning:
\begin{itemize}
    \item two LEO satellites and multiple antennas.
    \item three LEO satellites and a single antenna.
\end{itemize}
With a single time slot and no frequency synchronization across the LEO satellites and the receiver, the following configurations allow for $3$D positioning:
\begin{itemize}
    \item two LEO satellites and multiple antennas.
    \item three LEO satellites and a single antenna.
\end{itemize}
With a single time slot and no time and frequency synchronization across the LEO satellites and the receiver, the following configurations allow for $3$D positioning:
\begin{itemize}
    \item two LEO satellites and multiple antennas.
\end{itemize}
These results focused on the combinations of the number of LEO satellites and receive antenna that allows for $3$D positioning with a single time slot with no offset, time offset, frequency offset, and both offsets. 


More configurations that allow for $3$D positioning are presented next. These configurations are the minimal configurations that allow the FIM in Fig. \ref{EFIM_Position_3D} to be positive definite. With a single LEO satellite and no time or frequency offset, the following configurations allow for $3$D positioning:
\begin{itemize}
    \item two time slots and a single antenna.
\end{itemize}
With a single LEO satellite and no time synchronization across the LEO satellites and the receiver, the following configurations allow for $3$D positioning:
\begin{itemize}
    \item two time slots and multiple antennas.
    \item three time slots and a single antenna.
\end{itemize}
With a single LEO satellite and no time and frequency synchronization across the LEO satellites and the receiver, the following configurations allow for $3$D positioning:
\begin{itemize}
    \item two time slots and multiple antennas.
    \item three time slots and a single antenna.
\end{itemize}
These results focused on the combinations of the number of LEO satellites and receive antenna that allows for $3$D positioning with a single LEO satellite with no offset, time offset, frequency offset, and both offsets. In generating this result, we assumed that there is an SNR of $-20 \text{ dB}$ in all links, a center frequency of $1 \text{ GHz}$, and an effective bandwidth of $100 \text{ MHz}$.
 \subsection*{Available Information for $3$D Velocity Estimation}
\begin{figure}[H]
\centering
        {\includegraphics[clip,trim=0.2cm 0.08cm .2cm 0.01cm,
        width=\linewidth]{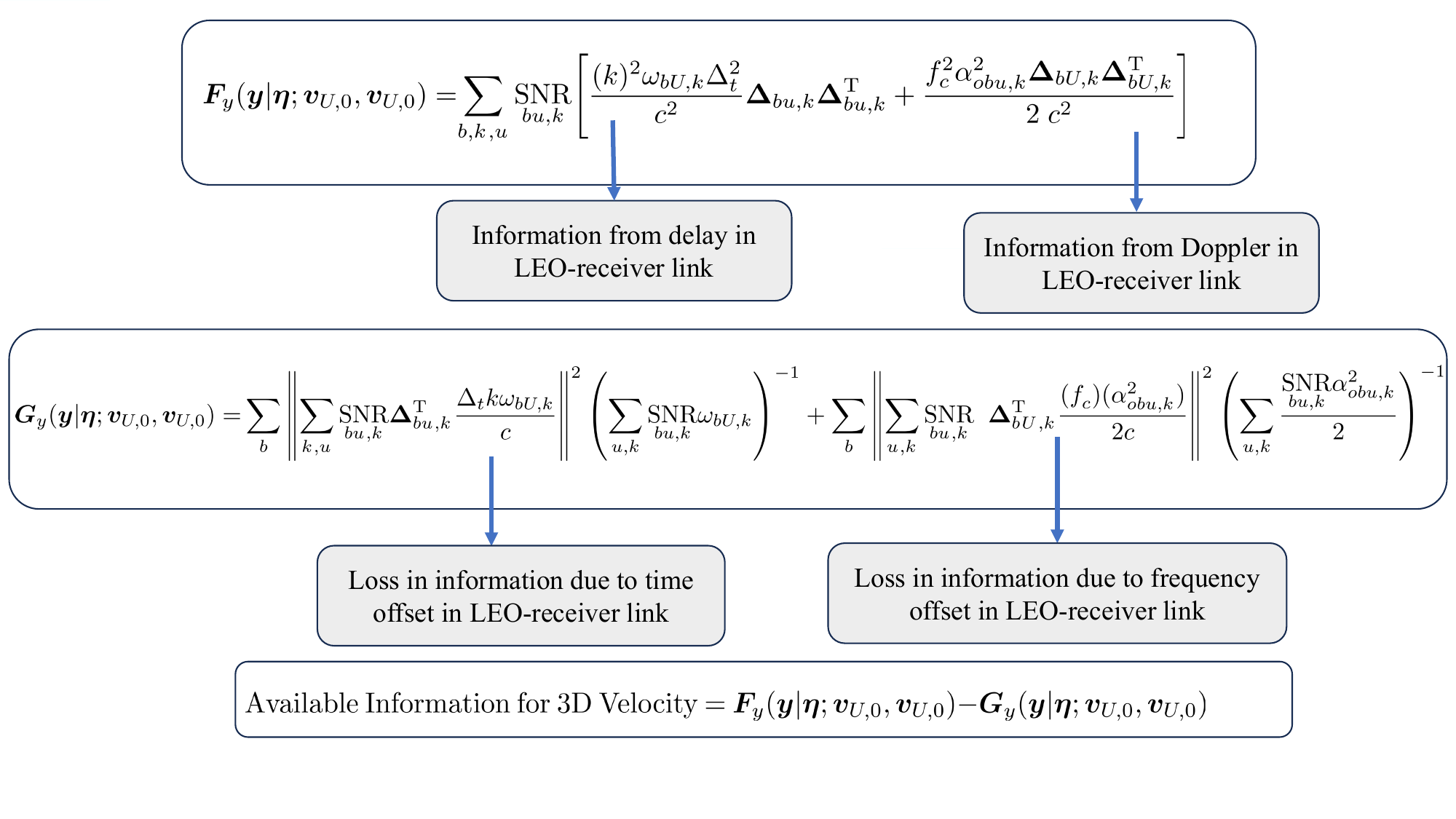}}
    \caption{The EFIM, $\mathbf{J}_{ \mathbf{\mathbf{y}}; \mathbf{\kappa}_1}^{\mathrm{e}}$, gives the available  information about $\mathbf{\kappa}_{1}$    present in $\mathbf{y}$. The entries in this EFIM, which are related to the $3$D velocity of the receiver, are given in this figure. This $3 \times 3$ matrix is located in the second block diagonal in $\mathbf{J}_{ \mathbf{\mathbf{y}}; \mathbf{\kappa}_1}^{\mathrm{e}}$.}
    \label{EFIM_Velocity_3D}
\end{figure}

 The FIM for the $3$D velocity estimation of the receiver is the second $3 \times 3$ matrix  on diagonal   of $\mathbf{J}_{ \bm{\bm{y}}; \bm{\kappa}_1}^{\mathrm{e}}$. This FIM is given by Fig. \ref{EFIM_Velocity_3D}.
The information from delays for estimating the $3$D velocity  of the receiver is 
$$
{{\sum_{b,k^{},u^{}} \underset{bu,k}{\operatorname{SNR}}   \frac{k^2 \omega_{bU,k}\Delta_{t}^{2}}{c^2}    \bm{\Delta}_{bu,k} 
\bm{\Delta}_{bu,k}^{\mathrm{T}} } }.
$$
This shows that the information provided by delay measurements for estimating the $3$D velocity of the receiver consists of the product of the SNR, the effective bandwidth, the time spacing between time slots, and the unit vectors from the LEO satellite to the receiver. Also, there is a sum across the number of LEO satellites, number of receive antennas, and the number of time slots, hence, the information from delays for estimating the $3$D velocity is a non-decreasing function of the number of LEO satellites, number of receive antennas, and the number of time slots. {\em From this equation, we notice that the information from delays from LEO satellites is significantly more than that from GPS satellites because the LEO satellites move faster than the GPS satellites. This is because, between time slots, the unit vector changes significantly more in the LEO-based system than in GPS systems. Hence, the FIM is increased significantly across time slots. This explains why Doppler based velocity estimation is primarily used in GPS systems, but in LEO-based systems, there exists a very real possibility of using both delays and Dopplers for velocity estimation.}
The information from Dopplers for estimating the $3$D velocity of the receiver is
$$
\sum_{b,k^{},u^{}} \underset{bu,k}{\operatorname{SNR}}   {\frac{f_{c}^2 \alpha_{obu,k}^2 \bm{\Delta}_{bU,k} \bm{\Delta}_{bU,k}^{\mathrm{T}}}{2 \; c^{2}}}. 
$$
Here, the information is a function of the product of the SNR, the transmit center frequency, and  root mean squared time duration; normalized by the squared of the speed of light. The loss in information due to the fact that the LEO satellites are not synchronized in time with each other and the receiver is 
$$
\sum_{b}   \norm{\sum_{k^{},u^{}} \underset{bu^{},k^{}}{\operatorname{SNR}}\bm{\Delta}_{bu^{},k^{}}^{\mathrm{T}} \frac{\Delta_{t}k \omega_{bU,k}}{c}}^{2}       \left(\sum_{u,k} \underset{bu,k}{\operatorname{SNR}} \omega_{bU,k}\right)^{\mathrm{-1}} .
$$
While the loss in information due to the fact that the LEO satellites are not synchronized in frequency with each other and the receiver is 
$$
\sum_{b}\norm{{\sum_{u^{},k^{} } \underset{bu^{},k^{}}{\operatorname{SNR}} \; \; \bm{\Delta}_{b^{}U^{},k^{}}^{\mathrm{T}}      }  \frac{(f_{c}^{}) (\alpha_{obu,k^{}}^{2})}{2 c} }^{2}  \left(\sum_{u,k} \frac{\underset{bu,k}{\operatorname{SNR}}  \alpha_{obu,k}^{2}}{2}\right)^{-1}.
$$
These errors scale linearly with an increase in the number of LEO satellites. {\em The structure of this loss in information is fundamentally different from the structure of the loss in information of GPS systems. This is because the GPS satellites are synchronized amongst each other in time and frequency, hence, the loss of information due to lack of synchronization between the GPS satellites and the receiver decreases much more slowly.} The structure of the loss of information due to synchronization errors in LEO satellites and GPS satellites is shown in (\ref{time_offset_loss_information_velocity} and  \ref{freq_offset_loss_information_velocity}).

\begin{equation}
\label{time_offset_loss_information_velocity}
    \begin{aligned}
&\underbrace{{{\frac{\Delta_{t}^{2}}{c^2} \sum_{b}} {  \norm{\sum_{k^{},u^{}} \underset{bu^{},k^{}} {\operatorname{SNR}} \; k \;    \bm{\Delta}_{bu,k}^{\mathrm{T}}  \omega_{bU,k} }^2
}    {   \left(\sum_{u,k} \underset{bu,k}{\operatorname{SNR}} \omega_{bU,k}\right)^{\mathrm{-1}}  } }
}_\text{loss in information due to lack of time synchronization among LEO satellites}   \succeq
\\ &\underbrace{{{\frac{\Delta_{t}^{2}}{c^2} } {  \norm{\sum_{b,k^{},u^{}} \underset{bu^{},k^{}} {\operatorname{SNR}} \; k \;    \bm{\Delta}_{bu,k}^{\mathrm{T}}  \omega_{bU,k} }^2
}    {   \left(\sum_{b,u,k} \underset{bu,k}{\operatorname{SNR}} \omega_{bU,k}\right)^{\mathrm{-1}}  } }
}_\text{loss in information due to lack of time. syn. between GPS satellites and the receiver} 
\end{aligned}
\end{equation}

\begin{equation}
\label{freq_offset_loss_information_velocity}
    \begin{aligned}
&\underbrace{\frac{1}{c^2}{\sum_{b}\norm{{\sum_{k^{},u^{} } \underset{bu^{},k^{}}{\operatorname{SNR}} \; \; \bm{\Delta}_{bU,k}^{\mathrm{T}}   }  \frac{(f_{c}^{}) (\alpha_{obu,k^{}}^{2})}{2} }^{2}  \left(\sum_{u,k} \frac{\underset{bu,k}{\operatorname{SNR}}  \alpha_{obu,k}^{2}}{2}\right)^{-1}}
}_\text{loss in information due to lack of freq. synchronization among LEO satellites}   \succeq
\\ &\underbrace{\frac{1}{c^2}{\norm{{\sum_{b,k^{},u^{} } \underset{bu^{},k^{}}{\operatorname{SNR}} \; \; \bm{\Delta}_{bU,k}^{\mathrm{T}}   }  \frac{(f_{c}^{}) (\alpha_{obu,k^{}}^{2})}{2} }^{2}  \left(\sum_{b,u,k} \frac{\underset{bu,k}{\operatorname{SNR}}  \alpha_{obu,k}^{2}}{2}\right)^{-1}}}_\text{loss in information due to lack of freq. syn. between GPS satellites and the receiver} 
\end{aligned}
\end{equation}


Equations \ref{time_offset_loss_information_velocity} and \ref{freq_offset_loss_information_velocity} shows the difference in the loss of information due to time and frequency offset when LEO satellites are used compared to GPS-based systems. The minimal configurations that allow for $3$D velocity estimation are presented next. These configurations are the minimal configurations that allow the FIM in Fig. \ref{EFIM_Velocity_3D} to be positive definite. With no time or frequency offset, the following configurations allow for $3$D velocity estimation:
\begin{itemize}
    \item one time slot, three LEO satellites, and a single antenna.
    \item two time slots, two LEO satellites, and a single antenna.
        \item three time slots, one LEO satellite, and a single antenna.
        \item four time slots, one LEO satellite, and a single antenna.
\end{itemize}
With time synchronization across the LEO satellites and the receiver, the following configurations allow for $3$D velocity estimation:
\begin{itemize}
    \item one time slot, three LEO satellites, and a single antenna.
    \item two time slots, two LEO satellites, and a single antenna.
        \item three time slots, one LEO satellite, and a single antenna.
        \item four time slots, one LEO satellite, and a single antenna.
\end{itemize}
With time and frequency synchronization across the LEO satellites and the receiver, the following configurations allow for $3$D velocity estimation:
\begin{itemize}
    \item two time slots, two LEO satellites, and a single antenna.
        \item three time slots, one LEO satellite, and a single antenna.
        \item four time slots, one LEO satellite, and a single antenna.
\end{itemize}
These results focused on the combinations of the number of LEO satellites and receive antenna that allows for $3$D velocity estimation with no offset, time offset, frequency offset, and both offsets. In generating this result, we assumed that there is an SNR of $-20 \text{ dB}$ in all links, a center frequency of $1 \text{ GHz}$, and an effective bandwidth of $100 \text{ MHz}$.


 \subsection*{Available Information for $3$D Orientation Estimation}
\begin{figure}[H]
\centering
    {\includegraphics[clip,trim=0.2cm 0.5cm .01cm 0.5cm,width=\linewidth]{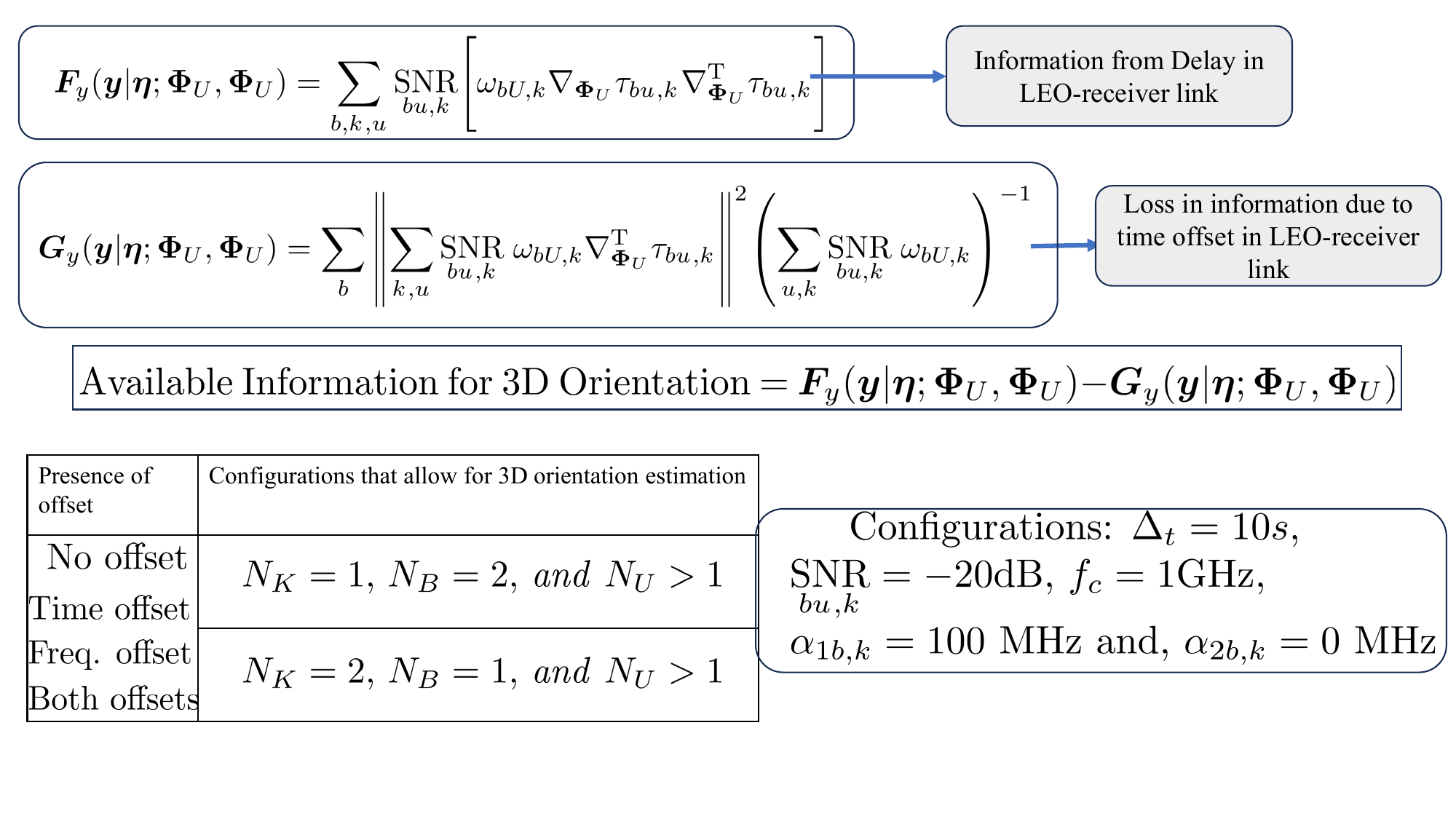}}
    \caption{This figure shows the information available from delays for $3$D orientation estimation and the loss in information due to time offset. This figure also shows the combinations of the number of LEO satellites, $N_B$, number of time slots, $N_K$, and receive antennas, $N_U$ that allows for $3$D orientation estimation  with no offset, time offset, frequency offset, and both offsets. In generating the combination results, we assumed that there is an SNR of $-20 \text{ dB}$ in all links, a time spacing of $\Delta_t = 10 \text{ s}$, a center frequency of $1 \text{ GHz}$, and an effective bandwidth of $100 \text{ MHz}$.} 
    \label{EFIM_Orientation_3D}
\end{figure}

 The FIM for the $3$D orientation estimation of the receiver is the third $3 \times 3$ matrix  on the diagonal   of $\mathbf{J}_{ \bm{\bm{y}}; \bm{\kappa}_1}^{\mathrm{e}}$. This FIM is given by Fig. \ref{EFIM_Orientation_3D}.
From Fig. \ref{EFIM_Orientation_3D}, we notice that the Dopplers are not useful for estimating the $3$D orientation, and the information from the delays for estimating the $3$D orientation is
$$
{\sum_{b,k^{},u^{}} \underset{bu,k}{\operatorname{SNR}}  \Bigg[\omega_{bU,k}  \nabla_{\bm{\Phi}_{U}}\tau_{bu^{},k^{}} 
 \nabla_{\bm{\Phi}_{U}}^{\mathrm{T}} \tau_{bu^{},k^{}}}    \Bigg].
$$
This shows that the orientation accuracy increases as the bandwidth and SNR increases. We also notice that multiple receive antennas is a necessary condition for estimating the $3$D orientation.

\subsection*{Available Information for $9$D Localization}
Our work primarily aims to provide the utility of using LEO satellites that are unsynchronized in time and frequency for $9$D receiver localization. The receiver $9$D localization can be estimated if $\mathbf{J}_{ \bm{\bm{y}}; \bm{\kappa}_1}^{\mathrm{e}}$ is positive definite. The $3 \times 3$ block diagonals, which correspond to the available information for $3$D positioning, available information for $3$D velocity estimation, and available information for $3$D orientation estimation, have been presented. The remaining block entries in $\mathbf{J}_{ \bm{\bm{y}}; \bm{\kappa}_1}^{\mathrm{e}}$ are the FIM relating the $3$D position and $3$D velocity, the FIM relating the $3$D position and $3$D orientation, and the FIM relating the $3$D velocity and $3$D orientation. These remaining entries can be found in \cite{Fundamentals_of_LEO_Based_Localization,Joint_9D_Receiver_Localization,emenonye2023_MILCOM_conf_9D_localization,emenonye2023_MILCOM_conf_9D_localization_1,emenonye2023_VTC_conf_Minimal,emenonye2023_VTC_conf_unsyn}.

The minimal configurations that allow for $9$D location estimation when there is both a lack of time and frequency synchronization between the LEO satellites and the receiver is three LEO satellites observed over three time slots with multiple receive antennas. These configurations are the minimal configurations that allow the FIM in the $9$D case to be positive definite. In generating this result, we assumed that there is an SNR of $-20 \text{ dB}$ in all links, a time spacing of $\Delta_t = 10 \text{ s}$, a center frequency of $1 \text{ GHz}$, and an effective bandwidth of $100 \text{ MHz}$.

\begin{figure*}
\centering
    {\includegraphics[clip,trim=0.2cm 1.5cm .2cm 0.1cm,width=\linewidth]{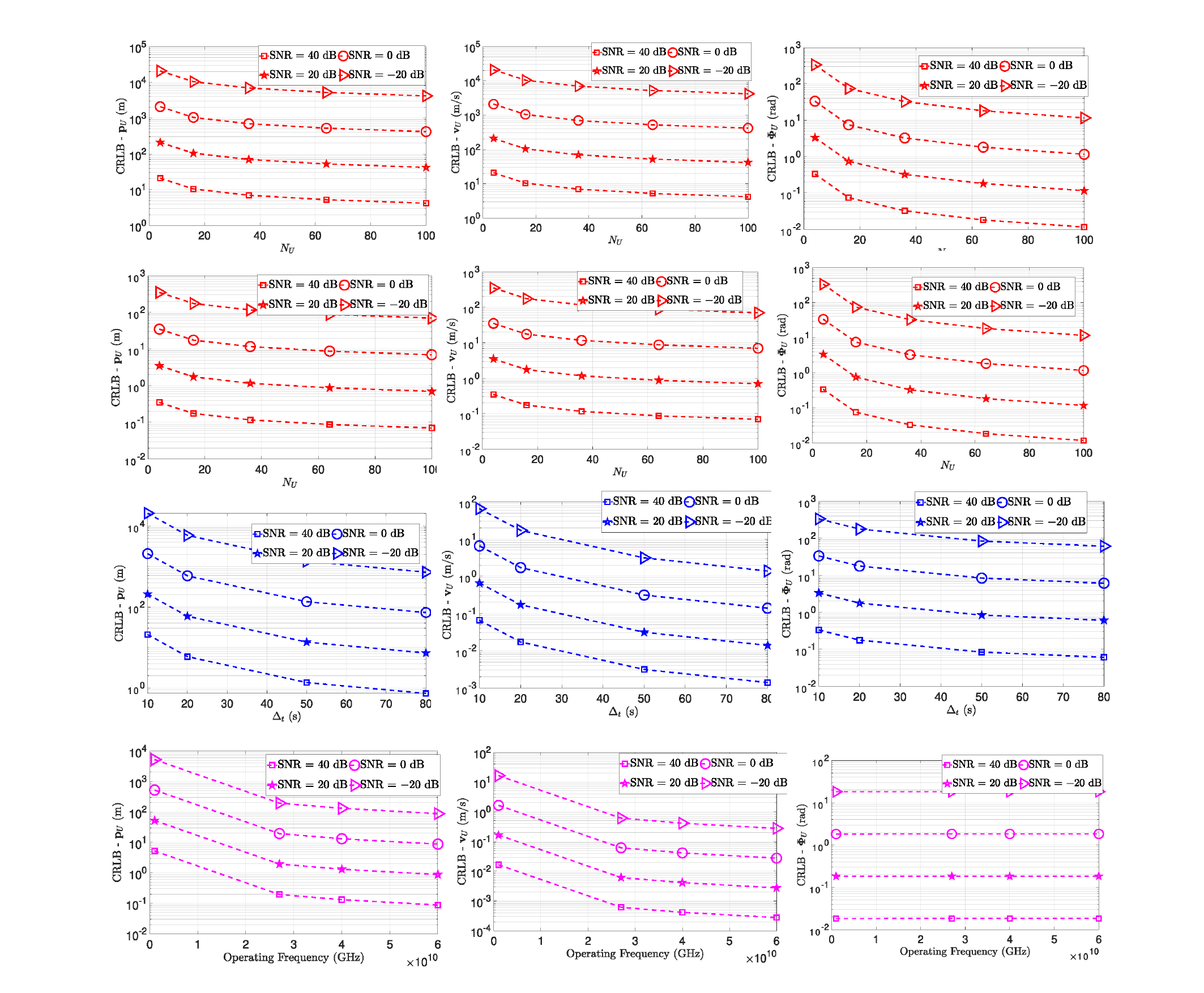}}
    \caption{The first column represents the CRLB for the $3$D positioning ($\bm{p}_U = \bm{p}_{U,0}$) as a function of the number of receive antenna, spacing between time slots, and the operating frequency. The second column represents the CRLB for the $3$D velocity estimation ($\bm{v}_U = \bm{v}_{U,0}$) as a function of the number of receive antenna, spacing between time slots, and the operating frequency. The third column represents the CRLB for the $3$D orientation estimation ($\bm{\Phi}_{U}$) as a function of the number of receive antenna, spacing between time slots, and the operating frequency.}
    \label{simulation_results}
\end{figure*}
Fig. 14 provides the CRLB of the positioning error obtained by inverting $\mathbf{J}_{ \bm{\bm{y}}; \bm{\kappa}_1}^{\mathrm{e}}$ and summing the first three diagonals, the CRLB of the velocity estimation error is obtained by inverting $\mathbf{J}_{ \bm{\bm{y}}; \bm{\kappa}_1}^{\mathrm{e}}$ and summing the following three diagonals, and the CRLB of the orientation estimation error is obtained by inverting $\mathbf{J}_{ \bm{\bm{y}}; \bm{\kappa}_1}^{\mathrm{e}}$ and summing the last three diagonals. For the CRLB of the positioning error, the first figure is obtained by considering a center frequency of $ 1 { \text{ GHz}}$ and a spacing of  $10\text{ s}$ between time slots, the curve representing an SNR of $-20 \text{ dB}$ shows that with four receive antennas a positioning accuracy of $20 \text{ km}$ is achievable, but with  100 receive antennas, this accuracy increases to $6 \text{ km}$. The curve representing an SNR of $0 \text{ dB}$ shows that with four receive antennas, a positioning accuracy of $2 \text{ km}$ is achievable, but with  100 receive antennas, this accuracy increases to $0.5 \text{ m}$. With an SNR of  $20 \text{ dB}$, the positioning accuracy increases from $200 \text{ m}$ with four receive antennas to $80 \text{ m}$ with hundred antennas. With an SNR of  $40 \text{ dB}$, the positioning accuracy increases from $20 \text{ m}$ with four receive antennas to $5 \text{ m}$ with hundred antennas.

The second figure (CRLB for positioning) is obtained by considering a center frequency of $ 60 { \text{ GHz}}$ and a spacing of  $10\text{ s}$ between time slots, the curve representing an SNR of $-20 \text{ dB}$ shows that as the number of receive antennas increases from four to a hundred antennas, the achievable positioning accuracy increases from $300 \text{ m}$ to $80 \text{ m}$. For an SNR of $0 \text{ dB}$, as the number of receive antennas increases from four to a hundred antennas, the achievable positioning accuracy increases from $30 \text{ m}$ to $8 \text{ m}$. With an SNR of  $20 \text{ dB}$, the positioning accuracy increases from $2 \text{ m}$ with four receive antennas to $0.9 \text{ m}$ with a hundred antennas. With an SNR of  $40 \text{ dB}$, the positioning accuracy increases from $0.3 \text{ m}$ with four receive antennas to $0.08 \text{ m}$ with a hundred antennas.

The third figure focusing on positioning is obtained by considering a center frequency of $ 1 { \text{ GHz}}$ and four receive antennas, and the spacing between time slots is varied. The curve representing an SNR of  $-20 \text{ dB}$ shows that the achievable accuracy increases from $20 \text{ km}$ to $0.8 \text{ km}$ as the spacing between time slots is varied from $10 \text{ s}$ to $80 \text{ s}$. While, with an SNR of $0 \text{ dB}$, the achievable accuracy increases from $3 \text{ km}$ to $0.1 \text{ km}$ as the spacing between time slots is varied from $10 \text{ s}$ to $80 \text{ s}$, the achievable accuracy considering $20 \text{ dB}$ increases from $0.2 \text{ km}$ to $10 \text{ m}$. With an SNR of $40 \text{ dB}$, the achievable accuracy increases from $20 \text{ m}$ to $0.9 \text{ m}$ as the spacing between time slots is varied from $10 \text{ s}$ to $80 \text{ s}$.

The fourth figure presents positioning accuracy as a function of the operating frequency with sixty-four receive antennas and a spacing of $10 \text{ s}$. The curve representing an SNR of  $-20 \text{ dB}$ shows that the achievable accuracy increases from $10 \text{ km}$ to $0.1 \text{ km}$ as the operating frequency is varied from $1 \text{ GHz}$ to $60 \text{ GHz}$. While, with an SNR of $0 \text{ dB}$, the achievable accuracy increases from $0.6 \text{ km}$ to $10 \text{ m}$  as the operating frequency is varied from $1 \text{ GHz}$ to $60 \text{ GHz}$, the achievable accuracy considering $20 \text{ dB}$ increases from $50 \text{ m}$ to $1 \text{ m}$. With an SNR of $40 \text{ dB}$, the achievable accuracy increases from $5 \text{ m}$ to $0.1 \text{ m}$  as the operating frequency is varied from $1 \text{ GHz}$ to $60 \text{ GHz}$. {\em These results show that we can achieve GPS-like positioning accuracy (cm-level) by observing three LEO satellites over three-time slots with a multiple antenna receiver.}

The CRLB of the velocity estimation error is shown in Fig.\ref{simulation_results}. In the first figure, for an SNR of $-20 \text{ dB}$, we notice the achieve velocity estimation error decreases from $20 \text{ km/s}$ to $6 \text{ km/s}$ as the number of receive antennas increases from four to hundred. For an SNR of $0 \text{ dB}$, we notice the achieve velocity estimation error decreases from $2 \text{ km/s}$ to $0.8 \text{ km/s}$ as the number of receive antennas increases from four to hundred. While for an SNR of $20 \text{ dB}$ and $40 \text{ dB}$, we notice the achieve velocity estimation error decreases from $0.2 \text{ km/s}$ to $0.08 \text{ km/s}$ and from $0.02 \text{ km/s}$ to $0.005 \text{ km/s}$, respectively. 

In the second figure, for an SNR of $-20 \text{ dB}$, we notice the achievable velocity estimation error decreases from $0.5 \text{ 
 km/s}$ to $80 \text{ m/s}$ as the number of receive antennas increases from four to hundred. For an SNR of $0 \text{ dB}$, we notice the achieve velocity estimation error decreases from $50 \text{ m/s}$ to $8 \text{ m/s}$ as the number of receive antennas increases from four to hundred. While for an SNR of $20 \text{ dB}$ and $40 \text{ dB}$, we notice the achieve velocity estimation error decreases from $5 \text{ m/s}$ to $1 \text{ m/s}$ and from $0.5 \text{ m/s}$ to $0.08 \text{ m/s}$, respectively. The third figure focuses on the velocity estimation error and is obtained by considering a center frequency of $ 1 { \text{ GHz}}$ and four receive antennas, and the spacing between time slots is varied. The curve representing an SNR of  $-20 \text{ dB}$ shows that the achievable accuracy increases from $100 \text{ m/s}$ to $1 \text{ m/s}$ as the spacing between time slots is varied from $10 \text{ s}$ to $80 \text{ s}$. While, with an SNR of $0 \text{ dB}$, the achievable accuracy increases from $10 \text{ m/s}$ to $0.1 \text{ m/s}$ as the spacing between time slots is varied from $10 \text{ s}$ to $80 \text{ s}$, the achievable accuracy considering $20 \text{ dB}$ increases from $1 \text{ m/s}$ to $0.01 \text{ m/s}$. With an SNR of $40 \text{ dB}$, the achievable accuracy increases from $0.1 \text{ m/s}$ to $0.001 \text{ m/s}$ as the spacing between time slots is varied from $10 \text{ s}$ to $80 \text{ s}$.

The fourth figure presents the velocity estimation accuracy as a function of the operating frequency with sixty-four receive antennas and a spacing of $10 \text{ s}$. The curve representing an SNR of  $-20 \text{ dB}$ shows that the achievable accuracy increases from $10 \text{ m/s}$ to $0.5 \text{ m/s}$ as the operating frequency is varied from $1 \text{ GHz}$ to $60 \text{ GHz}$. With an SNR of $0 \text{ dB}$, the achievable accuracy increases from $1 \text{ m/s}$ to $0.02 \text{ m/s}$  as the operating frequency is varied from $1 \text{ GHz}$ to $60 \text{ GHz}$. {\em These results show we can achieve GPS-like velocity estimation accuracy by observing three LEO satellites over three-time slots with a multiple antenna receiver.}

Focusing on the third column in Fig. \ref{simulation_results}, we notice that with some combinations of the number of receive antennas and spacing within time slots, we can achieve orientation estimation accuracy on the order of $10^{-2}$ rad. {\em It is essential to note that our analysis indicates that the operating frequency has no impact on the orientation estimation error, and this is a fundamental difference between orientation estimation and positioning where the operating frequency affects the CRLB.}

\section*{Summary}
This work has motivated the need for an alternative to the current GNSS systems and argued for a complete $9$D localization, which includes $3$D positioning, $3$D velocity estimation, and $3$D orientation estimation. We presented a comprehensive summary of the current GNSS systems: GPS, Galileo, BDS, GLONASS, IRNSS, and QZSS. 

We introduced some basic estimation theory concepts and connected these concepts to model identifiability. Here, model identifiability consists of utilizing the FIM to determine the number of receive antennas, number of time slots, and number of LEO satellites (unsynchronized in time and frequency) that enable $3$D positioning, $3$D velocity estimation, and $3$D orientation estimation. We also compare the structure of the FIM in LEO-based localization systems to that of the FIM in GPS systems. Lastly, we utilize the FIM to determine the utility of unsynchronized LEO satellites to jointly estimate the $3$D position, $3$D velocity, and $3$D orientation. Our analysis indicates that we can perform this $ 9$D localization with three LEO satellites observed over three-time slots as long as the receiver has multiple antennas.
\section*{Acknowledgment}
\label{sec:bio}
Don-Roberts Emenonye, Harpreet. S. Dhillon, and R. Michael.  Buehrer are with Wireless@VT,  Bradley Department of Electrical and Computer Engineering, Virginia Tech,  Blacksburg,
VA, 24061, USA. Email. Their email addresses are listed in the author sequence as donroberts@vt.edu, hdhillon@vt.edu, rbuehrer@vt.edu. The support of the US National Science Foundation (Grants ECCS-2030215, CNS-1923807, and CNS-2107276) is gratefully acknowledged.
\subsection*{Authors}
\begin{IEEEbiographynophoto}{Don-Roberts Emenonye}
(~\IEEEmembership{Graduate Student Member,~IEEE})  received the B.Sc. degree in electrical and electronics engineering 
from the University of Lagos, Nigeria, in 2016, and the M.S. degree in electrical engineering from Virginia Tech in 2020. 
He is currently pursuing the Ph.D. degree with the Bradley Department of Electrical and Computer Engineering, Virginia Tech, USA. 
His research interests span the area of communication theory and wireless positioning.
\end{IEEEbiographynophoto}
\begin{IEEEbiographynophoto}{Harpreet S. Dhillon}
(~\IEEEmembership{Fellow,~IEEE}) received his B.Tech. from IIT Guwahati (2008), M.S. from Virginia Tech (2010), and Ph.D. from the University of Texas at Austin (2013). He is currently the W. Martin Johnson Professor of Engineering and the Associate Dean for Research and Innovation at Virginia Tech. His research focuses on communication theory, wireless networks, geolocation, and stochastic geometry. He has received numerous awards for his contributions to these topics, including the IEEE Leonard G. Abraham Prize and IEEE Heinrich Hertz Award. He has served as TPC Co-chair for IEEE WCNC 2022 and IEEE PIMRC 2024, and is currently serving on the Executive Editorial Committee for the IEEE Transactions on Wireless Communications.
\end{IEEEbiographynophoto}
\begin{IEEEbiographynophoto}{R. Michael Buehrer}
(~\IEEEmembership{Fellow,~IEEE}) received his Ph.D.
degree in electrical engineering and computer science
from Virginia Tech, Virginia, in 1996. Dr. Buehrer was named an IEEE Fellow in 2016 “for contributions to wideband signal processing in communications and geolocation.”  His current research interests include machine learning for wireless communications and radar, geolocation, position location networks, cognitive radio, cognitive radar, electronic warfare, dynamic spectrum sharing, communication theory, Multiple Input Multiple Output (MIMO) communications, spread spectrum, interference avoidance, and propagation modeling.  His work has been funded by the National Science Foundation, the Defense Advanced Research Projects Agency, Office of Naval Research, the Army Research Lab, the Air Force Research Lab and several industrial sponsors. 
\end{IEEEbiographynophoto}

{
\bibliographystyle{IEEEtran}
\bibliography{refs}
}

\end{document}